\documentclass[12pt]{article}
\pdfoutput=1
\usepackage{draft}
\usepackage{comment}
\usepackage[normalem]{ulem}
\usepackage{hyperref}
\hypersetup{
 colorlinks=true,
 urlcolor=blue,
 anchorcolor=blue,
 citecolor=blue,
 filecolor=blue,
 linkcolor=blue,
 menucolor=blue,
 pagecolor=blue,
 linktocpage=true,
}
\usepackage{graphicx,color,subfig}
\usepackage{cite}
\usepackage{empheq} 
\usepackage[font=small,labelfont=bf]{caption} 

\usepackage[usenames,dvipsnames,table]{xcolor}
\usepackage{tikz}
\usepackage{empheq}

\renewcommand{\Re}{\operatorname{Re}}

\usepackage{scalerel}
\usepackage{stackengine,wasysym}

\DeclareFontFamily{OT1}{pzc}{}
\DeclareFontShape{OT1}{pzc}{m}{it}{<-> s * [1.10] pzcmi7t}{}
\DeclareMathAlphabet{\mathpzc}{OT1}{pzc}{m}{it}

\definecolor{vert}{rgb}{0.1367 0.543 0.1367}

\def\({\left(}
\def\){\right)}
\newcommand{\nn}{\nonumber}

\begin{document}

\unitlength = .8mm

\begin{titlepage}

	  \begin{flushright}
	 \hfill{\tt CALT-TH 2022-026}
  \end{flushright} 

\begin{center}

 \hfill \\
 \hfill \\

\title{Scalar Modular Bootstrap and \\ Zeros of the Riemann Zeta Function}
% Scalar Modular Bootstrap and the Riemann Hypothesis
% Mathematicians hate them! Click here to see how local two physicists proved the Riemann Hypothesis
% The Scalar Modular Bootstrap
% Modular Bootstrap and Zeros of the Zeta Function
% High Temperature Limit of 2d CFT partition functions
% Harmonic analysis and modular bootstrap

~\vskip 0.01 in

\author{
	Nathan Benjamin and Cyuan-Han Chang
}
~\vskip 0.05 in

\emph{\small Walter Burke Institute for Theoretical Physics \\ California Institute of Technology, Pasadena, CA 91125, USA}
~\vskip .2 in
\email{
	nbenjami@caltech.edu, cchang7@caltech.edu
}

\end{center}

\abstract{
Using the technology of harmonic analysis, we derive a crossing equation that acts only on the scalar primary operators of any two-dimensional conformal field theory with $U(1)^c$ symmetry. From this crossing equation, we derive bounds on the scalar gap of all such theories. Rather remarkably, our crossing equation contains information about all nontrivial zeros of the Riemann zeta function. As a result, we rephrase the Riemann hypothesis purely as a statement about the asymptotic density of scalar operators in certain two-dimensional conformal field theories. We discuss generalizations to theories with only Virasoro symmetry. 
}

\vfill

\end{titlepage}

\eject

\begingroup

\baselineskip .168 in
\tableofcontents

\endgroup

\section{Introduction}
\label{sec:intro} 

The conformal bootstrap is a powerful program used to highly constrain quantum field theories starting from basic consistency conditions. In two dimensional conformal field theory (CFT), one avatar of this program is the so-called modular bootstrap which uses modular invariance of the genus one partition function to constrain possible allowed spectra of 2d CFTs. This program started with the work of \cite{Hellerman:2009bu} and has led to many interesting results (see e.g. \cite{Keller:2012mr,Friedan:2013cba, Benjamin:2016fhe, Collier:2016cls, Bae:2017kcl, Afkhami-Jeddi:2019zci, Mukhametzhanov:2019pzy, Hartman:2019pcd, Benjamin:2019stq, Pal:2019zzr, Alday:2019vdr, Afkhami-Jeddi:2020hde} for a non-exhaustive list). This has several applications, including constraining theories of quantum gravity in AdS$_3$.

In many (but not all) cases, the \emph{spinless} bootstrap equations are studied, in which one throws away information about the spin of the original operators and only looks at their energies. This is done by grading the partition function only by the energies of the operators, and using $S$-invariance, rather than the full $SL(2,\mathbb Z)$-invariance of the partition function. In particular, we have
\be
Z(y) \coloneqq \sum_{\mathcal{O}} e^{-2\pi y (\Delta_{\mathcal{O}}-\frac{c}{12})} = Z(y^{-1}),
\label{eq:modularS}
\ee
where the sum over $\mathcal{O}$ is a sum over all local operators in the theory, and $\Delta_\mathcal{O}$ is the scaling dimension of operator $\mathcal{O}$.
Any bound derived from (\ref{eq:modularS}) will by definition be insensitive to the spins of the operators $\mathcal{O}$. For example, the current strongest bound on the lightest nontrivial Virasoro primary operator at large central charge $c$ is in \cite{Afkhami-Jeddi:2019zci}, which showed at large $c$, 
\be
\Delta^{\text{Virasoro}}_{\text{gap}} \lesssim \frac{c}{9.1}.
\ee 
However it makes no claim on what the spin of that operator is, or what the lightest spin $j$ operator is. A similar result using the spinless bootstrap was found for a simpler class of theories, those with a $U(1)^c$ chiral algebra, in \cite{Afkhami-Jeddi:2020hde}
\be
\Delta^{U(1)^c}_{\text{gap}} \lesssim \frac{c}{9.869}.
\label{eq:u1gap}
\ee

In this paper we derive a novel one-dimensional crossing equation using the technology of harmonic analysis. In the case of CFTs with $U(1)^c$ symmetry, this crossing equation acts \emph{only} on the scalar primary operators of the theory (with respect to the $U(1)^c$ chiral algebra). This allows us to place new bounds on the scalar gap of all $U(1)^c$ conformal field theories for any integer $c$. This is more refined information than the bound in e.g. (\ref{eq:u1gap}) since it provides explicit information about the spin of the operator. Indeed the scalar gap is a natural object to consider. Scalar operators can be added to the Lagrangian while still preserving Lorentz invariance. The scalar gap is then related to questions about, for instance, if the CFT has a relevant operator or not. Another application is in the study of boundary conformal field theory. There, the bulk scalars show up in some crossing equations rather than all bulk operators, which can lead to interesting bounds that are conditional on the scalar gap \cite{Collier:2021ngi}. 

Remarkably, our crossing equation has an intimate relation with the nontrivial zeros of the Riemann zeta function. In a sense which we will explain, hidden inside the scalar operators of any 2d CFT with $U(1)^c$ symmetry are the nontrivial zeros of the zeta function. As a result, we can rephrase the Riemann hypothesis as a statement about the behavior of scalar operators of any $U(1)^c$ CFT. 

We also discuss a generalization to Virasoro CFTs. We derive a more complicated one-dimensional crossing equation that involves operators of all spins. The nontrivial zeros of the zeta function again play an important role. This leads to the Riemann hypothesis being equivalent to a more complicated statement about the asymptotic density of a signed count of all operators (of any spin) in any CFT. Unfortunately we run into some technical obstacles in bounding physical quantities such as the scalar gap for Virasoro CFTs.

This paper is organized as follows. In Section \ref{sec:review} we review harmonic analysis on the fundamental domain of $SL(2,\mathbb Z)$, which will play an important role in deriving our scalar crossing equation. In Section \ref{sec:narain} we apply this to the study of $U(1)^c$ CFTs and derive the scalar crossing equation. We present the numerical results for the scalar gap of $U(1)^c$ theories for various values of $c$. In Section \ref{sec:virasoro} we discuss generalizations to theories with only Virasoro symmetry. In Section \ref{sec:RH} we study more explicitly the connections between 2d CFTs and the Riemann hypothesis. We discuss various potentially interesting future directions in Section \ref{sec:conclusion}. Some detailed calculations and derivations are banished to the appendices.

\section{Review of Harmonic Analysis}
\label{sec:review}

In this section we will review harmonic analysis on the space $\mathbb{H}/SL(2,\mathbb Z)$, where $\mathbb{H}$ is the upper half plane. For much of this discussion, we refer to \cite{Terras_2013}. We will use the notation of \cite{Benjamin:2021ygh} in this section.

The main idea is to decompose square-integrable modular invariant functions into eigenfunctions of the Laplacian on the space $\mathbb{H}/SL(2,\mathbb Z)$. If $\tau \in \mathbb{H}$, with real and imaginary parts $x, y$ respectively, then there is a natural metric on $\mathbb{H}$ given by
\be
ds^2 = \frac{dx^2 + dy^2}{y^2}.
\label{eq:measure}
\ee
The Laplacian on this space is given by
\be
\Delta = -y^2(\partial_x^2 + \partial_y^2).
\label{eq:laplace}
\ee
Square-integrable modular-invariant functions $f(\tau)$ are those with finite $L^2$ norm under the measure (\ref{eq:measure}), meaning
\be
\int_{-1/2}^{1/2} dx \int_{\sqrt{1-x^2}}^\infty \frac{dy}{y^2} |f(\tau)|^2 < \infty.
\label{eq:l2}
\ee

If $f(\tau)$ is a square-integrable, modular-invariant function, it has a unique decomposition into eigenfunctions of the Laplacian (\ref{eq:laplace}). These eigenfunctions have been classified and they come in three types:
\begin{itemize}
    \item The constant function $1$, with eigenvalue $0$.
    \item An infinite, continuous family of eigenfunctions known as real analytic Eisenstein series, $E_s(\tau)$, with $s=\frac12 + i t$, $t$ real, with eigenvalue $\frac14 + t^2$. Any real $t$ is permissible.
    \item An infinite, discrete family of eigenfunctions known as Maass cups forms, denoted $\nu^\pm_n(\tau)$, $n=1, 2, \cdots$. These have sporadic eigenvalues, which we denote $\frac14 + (R^\pm_n)^2$, for $R^{\pm}_n$ a positive real number. Both $\nu^+_n$ and $\nu^-_n$ are ordered in increasing eigenvalue, i.e. $R_1^+ < R_2^+ < \cdots$, and likewise for $R_n^-$. The superscript $\pm$ refers to whether the cusp form is even or odd under parity.
\end{itemize}
The decomposition of $f(\tau)$ is then given by:
\be
f(\tau) = \frac{(f, 1)}{(1,1)} + \frac{1}{4\pi i}\int_{\frac12-i\infty}^{\frac12+i\infty} ds E_s(\tau) (f, E_s) + \sum_{n=1}^{\infty} \sum_{\epsilon = \pm} \nu^\epsilon_n(\tau) \frac{(f, \nu^\epsilon_n)}{(\nu^\epsilon_n,\nu^\epsilon_n)},
\label{eq:roeclke}
\ee
where the overlap function is given by the Petersson inner product:
\be
(f, g) \coloneqq \int_{-1/2}^{1/2} dx \int_{\sqrt{1-x^2}}^\infty \frac{dy}{y^2} f(\tau) \overline{g(\tau)}.
\ee
The decomposition (\ref{eq:roeclke}) is known as the Roelcke-Selberg decomposition.

Let us be more explicit about the eigenfunctions of the Laplacian. The real analytic Eisenstein series $E_s(\tau)$, $s\in \mathbb{C}$ are defined as a modular sum of $y^s$:
\be
E_s(\tau) = \sum_{\gamma \in \Gamma_{\infty}\backslash SL(2,\mathbb Z)} y^s|_{\gamma},
\label{eq:poincare}
\ee
where $\Gamma_\infty$ is the subgroup of $SL(2,\mathbb Z)$ generated by $\tau \rightarrow \tau+1$. The sum (\ref{eq:poincare}) converges if $\text{Re}(s) > 1$. However, it admits an analytic continuation everywhere in the $s$ plane:
\be
E_s(\tau) = y^s + \frac{\Lambda(1-s)}{\Lambda(s)} y^{1-s} + \sum_{j=1}^\infty \frac{4\sigma_{2s-1}(j)\sqrt{y} K_{s-\frac12}(2\pi j y)}{\Lambda(s) j^{s-\frac12}} \cos(2\pi j x),
\label{eq:estrue}
\ee
where $\sigma_{2s-1}(j)$ is the divisor sigma function, $K$ is the modified Bessel function of second kind, and $\Lambda$ is defined as
\be
\Lambda(s) \coloneqq \pi^{-s} \zeta(2s) \Gamma(s).
\label{eq:lambdadef}
\ee
The function $\Lambda(s)$ obeys a useful identity:
\be
\Lambda(s) = \Lambda(\tfrac12 - s).
\label{eq:Lambdafunceq}
\ee
From (\ref{eq:estrue}) we also see that the real analytic Eisenstein series obey a useful identity:
\be
\Lambda(s) E_s(\tau) = \Lambda(1-s) E_{1-s}(\tau).
\label{eq:useful}
\ee

The remaining eigenfunctions, the Maass cusp forms, are more mysterious. They take the following functional form:
\begin{align}
\nu^+_n(\tau) &= \sum_{j=1}^\infty a_j^{(n,+)} \sqrt y K_{i R_n^+}(2\pi j y) \cos(2\pi j x) \nn\\
\nu^-_n(\tau) &= \sum_{j=1}^\infty a_j^{(n,-)} \sqrt y K_{i R_n^-}(2\pi j y) \sin(2\pi j x),
\end{align}
where $R_n^\pm$ and $a_j^{(n,\pm)}$ are a set of sporadic real numbers. For example, we have the following first few values of $R_n^\pm$:
\begin{align}
    &R_1^+ \approx 13.77975,~~~~ R_1^- \approx 9.53370 \nn\\
    &R_2^+ \approx 17.73856, ~~~~ R_2^- \approx 12.17301 \nn\\
    &R_3^+ \approx 19.42348, ~~~~ R_3^- \approx 14.35851.
\end{align}
For more numerical data on the Maass cusp forms, see the online database \cite{LMFDB}. One key feature the Maass cusp forms have is, unlike the real analytic Eisenstein series, they all lack a scalar piece:
\be
\int_{-1/2}^{1/2} dx \nu_n^\pm(\tau) = 0.
\label{eq:massnoscalar}
\ee

\section{$U(1)^c$ CFTs}
\label{sec:narain}

We begin with studying a family of particularly simple conformal field theories, with an extended current algebra of $U(1)^c$. Examples of such CFTs include Narain's family of $c$ free bosons compactified on a $c$-dimensional lattice, parameterized by the moduli space $O(c,c,\mathbb{Z})\backslash O(c,c)/O(c) \times O(c)$. It is believed that this family of CFTs fully classifies all theories with $U(1)^c$ current algebra. However this has not been proven. Our results in this section will apply to all theories with $U(1)^c$ symmetry; we do not need to assume the theory is a Narain CFT.

\subsection{Harmonic decomposition}
In \cite{Benjamin:2021ygh}, the harmonic decomposition of $U(1)^c$ CFT partition functions were calculated, which we review here. The characters of the $U(1)^c$ chiral algebra are given by
\be
\chi^h(\tau) = \frac{q^h}{\eta(\tau)^c},
\ee
where $\eta(\tau)$ is the Dedekind eta function. Instead of decomposing the full partition function $Z(\tau)$, we instead consider the primary-counting partition function
\begin{align}
\hat{Z}^c(\tau, \mu) &\coloneqq y^{c/2} |\eta(\tau)|^{2c} Z(\tau) \nn\\
&= y^{c/2} \sum_{h,\bar{h}} q^h \bar{q}^h,
\label{eq:primarycounting}
\end{align}
where in (\ref{eq:primarycounting}) the sum over $h, \bar h$ goes over the $U(1)^c$ primary operators. In (\ref{eq:primarycounting}), we write $\hat{Z}^c(\tau, \mu)$ to emphasize that the (reduced) partition function depends not only on the worldsheet modulus $\tau$, but also on an abstract target space coordinate $\mu$.\footnote{For Narain theories, we can view $\mu$ as a parameter $\mu \in O(c,c;\mathbb{Z}) \backslash O(c,c) / O(c) \times O(c)$. The target space of Narain theories is parametrized by a symmetric metric $G_{ab}$ and an antisymmetric $B$-field $B_{ab}$, where $a,b$ indices run from $1, 2, \cdots, c$. Here, however, we can just view $\mu$ as some abstract coordinate.}

The function (\ref{eq:primarycounting}) is not yet square-integrable, but once we subtract out the Eisenstein series $E_{c/2}(\tau)$ (defined in (\ref{eq:estrue})), this yields a square-integrable function that admits a unique spectral decomposition\footnote{For Narain CFTs, $E_{c/2}(\tau)$ has the interpretation of the averaged partition function \cite{Afkhami-Jeddi:2020ezh, Maloney:2020nni}.}. In \cite{Obers:1999um, Angelantonj:2011br, Benjamin:2021ygh} the spectral decomposition was given as follows\footnote{Note that due to the pole structure of $\Lambda(s)$ and the real analytic Eisenstein series $E_s(\tau)$, the decompositions of $c=1$ and $c=2$ are slightly different than other $c$, so we will assume $c\neq 1, 2$ for the rest of this section. We revisit $c=1$ and $c=2$ in Appendix \ref{sec:c1}.}:
\begin{align}
    \hat{Z}^c(\tau, \mu) &= E_{c/2}(\tau) + 3\pi^{-\frac c2} \Gamma\left(\frac c2 -1\right) \mathcal{E}^c_{\frac c2-1}(\mu) + \frac1{4\pi i}\int_{\frac12-i\infty}^{\frac12+i\infty} ds \pi^{s-\frac c2} \Gamma\left(\frac c2-s\right)\mathcal{E}_{\frac c2-s}^c(\mu) E_s(\tau) \nn\\
    &+ \sum_{n=1}^\infty \sum_{\epsilon = \pm} \frac{(\hat{Z}^c, \nu_n^\epsilon)(\mu)}{(\nu_n^\epsilon, \nu_n^\epsilon)} \nu^\epsilon_n(\tau).
    \label{eq:spectraldecomp}
\end{align}
The coefficients $\mathcal{E}_s^c(\mu)$ were called constrained Epstein zeta series in \cite{Angelantonj:2011br}, and are defined as:
\be
\mathcal{E}_s^c(\mu) \coloneqq \sum_{\Delta\in\mathcal{S}} (2\Delta)^{-s},
\label{eq:oddeis}
\ee
where we define the set $\mathcal{S}$ to be the dimensions of all non-vacuum scalar primary operators under the $U(1)^c$ chiral algebra (with multiplicity). This sum converges for $\text{Re}(s) > c-1$, but like for the $SL(2,\mathbb Z)$ Eisenstein series (\ref{eq:poincare}), they admit an analytic continuation everywhere in the complex $s$ plane. They also obey a functional equation:
\be
\mathcal{E}^c_{\frac c2 - s}(\mu) = \frac{\Gamma(s)\Gamma(s+\frac c2-1)\zeta(2s)}{\pi^{2s-\frac12}\Gamma(\frac c2-s)\Gamma(s-\frac 12)\zeta(2s-1)} \mathcal{E}^c_{\frac c2 + s-1}(\mu).
\label{eq:varepsfunc}
\ee
This equation is inherited from the functional equation that the Eisenstein series obey (\ref{eq:useful}), combined with the definition of $\mathcal{E}^c_s(\mu)$ as an overlap of $\hat{Z}^c(\tau, \mu)$ with the Eisenstein series:
\be
(\hat{Z}^c-E_{\frac c2},E_s) = \pi^{s-\frac c2}\Gamma\left(\frac c2 - s\right) \mathcal{E}^c_{\frac c2-s}(\mu).
\ee

For Narain CFTs, (\ref{eq:oddeis}) can be rewritten as
\be
\mathcal{E}_s^c(\mu) = \sideset{}{'}\sum_{\vec{n}, \vec{m}\in \mathbb{Z}^c} \frac{\delta_{\vec{n}\cdot \vec{w},0}}{M_{\vec{n},\vec{w}}(\mu)^{2s}},
\label{eq:oddeis2}
\ee
with
\be
M_{\vec{n},\vec{w}}(\mu)^2 \coloneqq G^{ab}(n_a + B_{ac}w^c)(n_b+B_{bd}w^d) + G_{cd} w^c w^d,
\label{eq:mdef}
\ee
and the prime over the summation indicating we should not sum over the vacuum state (with $\vec{n} = \vec{w} = \vec{0}$).

\subsection{Crossing equation}
\label{sec:cross}

Since the Maass cusp forms have no scalar piece (i.e. (\ref{eq:massnoscalar})), the scalar part of (\ref{eq:spectraldecomp}) is particularly simple:
\begin{align}
    \int_{-1/2}^{1/2} dx \hat{Z}^c(\tau, \mu) &= y^{\frac c2} + \frac{\Lambda\left(\frac{c-1}2\right)}{\Lambda\left(\frac c2\right)} y^{1-\frac c2} + 3\pi^{-\frac c2} \Gamma\left(\frac c2 -1\right) \mathcal{E}^c_{\frac c2-1}(\mu) \nn\\&+ \frac1{4\pi i} \int_{\frac12-i\infty}^{\frac12 + i \infty} ds \pi^{s-\frac c2}\Gamma\left(\frac c2-s\right) \mathcal{E}_{\frac c2-s}^c(\mu)\left(y^s + \frac{\Lambda(1-s)}{\Lambda(s)} y^{1-s}\right),
    \label{eq:firstscalar}
\end{align}
where as usual $\tau=x+iy$, and $\Lambda(s)$ is defined as in (\ref{eq:lambdadef}).

As a reminder, the set $\mathcal{S}$ is the set of conformal weights of all non-vacuum scalar primaries under the $U(1)^c$ chiral algebra (with multiplicity). We can rewrite the LHS of (\ref{eq:firstscalar}) as 
\be
\int_{-1/2}^{1/2} dx \hat{Z}^c(\tau, \mu) = y^{\frac c2} \left(1+\sum_{\Delta \in \mathcal{S}} e^{-2\pi \Delta y}\right).
\label{eq:scalarpartLHS}
\ee
This gives
\begin{align}
    \sum_{\Delta \in \mathcal{S}} e^{-2\pi \Delta y} &= \frac{\Lambda\left(\frac{c-1}2\right)}{\Lambda\left(\frac c2\right)} y^{1-c} + \varepsilon_c(\mu) y^{-\frac c2} + \frac1{2\pi i} \int_{\frac12-i\infty}^{\frac12+i\infty} ds \pi^{s-\frac c2} \Gamma\left(\frac c2-s\right)\mathcal{E}_{\frac c2-s}^c(\mu)y^{s-\frac c2},
    \label{eq:scalarsnotyetint}
\end{align}
where we have defined $\varepsilon_c(\mu) \coloneqq 3\pi^{-\frac c2}\Gamma\left(\frac c2-1\right)\mathcal{E}_{\frac c2-1}^c(\mu)$, and used the symmetry between $s \leftrightarrow 1-s$ in the integral over $s$.

The remaining task is to do the integral in (\ref{eq:scalarsnotyetint}). We will do the integral over $s$ by moving the contour to the right of $s=\frac c2$. It turns out the only poles we enclose after moving the contour are at $s=\frac c2, \frac{1+z_n}2, \frac{1+z_n^*}2$, where $z_n$ are the nontrivial zeros of the Riemann zeta function with positive imaginary part (i.e. $z_1 \approx \frac12 + 14.135i,~ z_2 \approx \frac12 + 21.022i$, etc.). See Fig. \ref{fig:polesc2} for a picture of the pole structure (shown for $c=3$). We derive the pole structure in Appendix \ref{sec:crossder}. After moving the contour, (\ref{eq:scalarsnotyetint}) becomes
\begin{align}
1 + \sum_{\Delta \in \mathcal{S}} e^{-2\pi \Delta y} &=  \frac{\Lambda\left(\frac{c-1}2\right)}{\Lambda\left(\frac c2\right)} y^{1-c} + \varepsilon_c(\mu) y^{-\frac{c}2} + \sum_{k=1}^\infty \text{Re}\left(\delta_{k,c}(\mu)y^{-\frac c2 + 1 - \frac{z_k}2}\right) \nonumber \\
&+ \frac1{2\pi i} \int_{\gamma-i\infty}^{\gamma+i\infty} ds \pi^{s-\frac c2} \Gamma\left(\frac c2-s\right)\mathcal{E}_{\frac c2-s}^c(\mu)y^{s-\frac c2},
\label{eq:scalarsalmostthere}
\end{align}
where $\gamma>\frac c2$. The terms $\varepsilon_c(\mu)$ and $\delta_{k,c}(\mu)$ are moduli-dependent constants, which have an explicit formula as
\begin{align}
 \varepsilon_c(\mu) &= \frac3\pi \int_{\mathcal{F}} \frac{dx dy}{y^2} (\hat{Z}^c(\tau,\mu) - E_{c/2}(\tau)) \nonumber \\
 \delta_{k,c}(\mu) &= \int_{\mathcal{F}} \frac{dx dy}{y^2} (\hat{Z}^c(\tau, \mu) - E_{c/2}(\tau)) \text{Res}_{s=z_k/2}E_{s}(\tau),
 \label{eq:residuedelta}
\end{align}
where
\be
\text{Res}_{s=z_k/2}E_{s}(\tau) = \frac{\sqrt\pi \zeta(z_k-1)\Gamma(\frac{z_k-1}2)}{2\zeta'(z_k)\Gamma(\frac{z_k}2)}y^{1-\frac{z_k}2} + \sum_{j=1}^\infty \frac{2\pi^{\frac{z_k}2}\cos(2\pi j x)\sigma_{z_k-1}(j)\sqrt y K_{\frac{z_k-1}{2}}(2\pi j y)}{j^{\frac{z_k-1}2}\zeta'(z_k)\Gamma(\frac{z_k}2)}.
\ee

\begin{figure}
    \centering
    \subfloat[\centering ]{{\includegraphics[width=7.25cm]{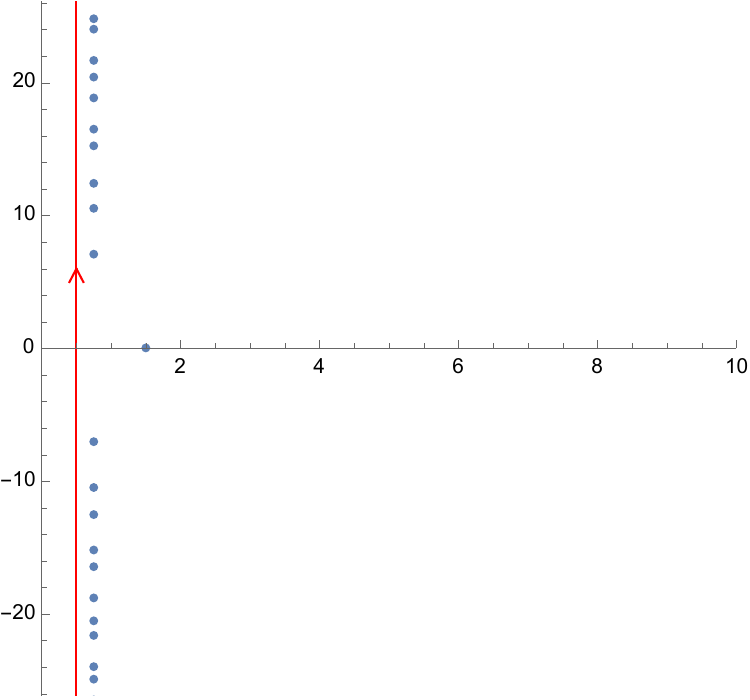} }}
    \qquad
    \subfloat[\centering ]{{\includegraphics[width=7.25cm]{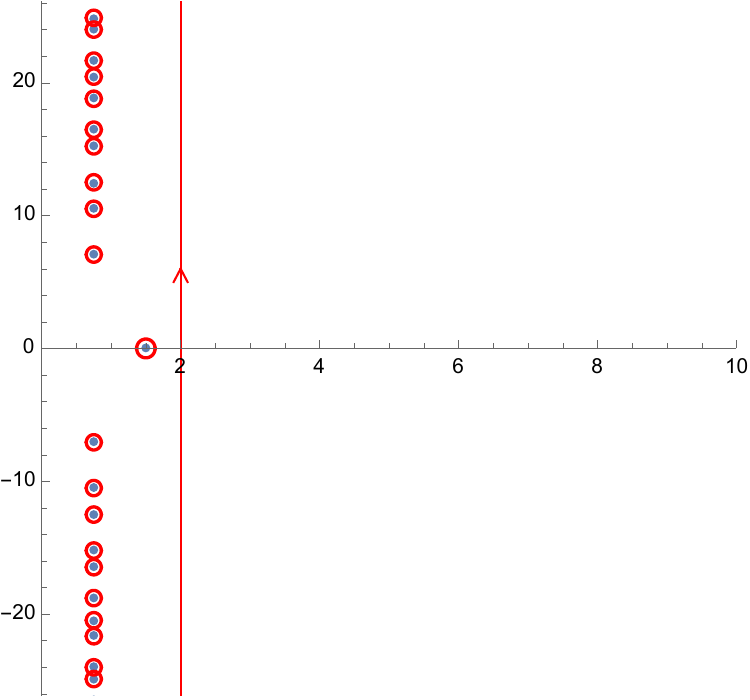} }} 
    \caption{{\bf(a)} Pole structure of the integral in (\ref{eq:scalarsnotyetint}) in the complex $s$ plane. The poles are located at $s=\frac c2, \frac{1+z_n}2, \frac{1+z_n^*}2$ (shown here for $c=3$), where $z_n$ are the nontrivial zeros of the Riemann zeta function with positive imaginary part. If the Riemann hypothesis is true, the tower of poles in the figure all occur at real part $\frac34$, except for the pole at $s=\frac c2$. {\bf (b)} Contour deformation of the integral to $\text{Re}(s) > \frac c2$.}
    \label{fig:polesc2}
\end{figure}

Now let us consider the integral in (\ref{eq:scalarsalmostthere}). We first rewrite the integral using the functional identity (\ref{eq:varepsfunc}):
\be
\int_{\gamma-i\infty}^{\gamma+i\infty} ds \pi^{s-\frac c2} \Gamma\left(\frac c2-s\right)\mathcal{E}_{\frac c2-s}^c(\mu)y^{s-\frac c2} = \int_{\gamma-i\infty}^{\gamma+i\infty} ds \frac{\Gamma(s)\Gamma(s+\frac c2-1)\zeta(2s)}{\pi^{s+\frac{c-1}2}\Gamma(s-\frac 12)\zeta(2s-1)} \mathcal{E}^c_{\frac c2 + s-1}(\mu) y^{s-\frac c2}.
\label{eq:gammastuff}
\ee
Because we take $\gamma > \frac c2$, this means that $\text{Re}(\frac c2 + s -1) > c-1$, which means we can write this as the following convergent sum:
\be
\mathcal{E}^c_{\frac c2+s-1}(\mu) = \sum_{\Delta \in \mathcal{S}} (2\Delta)^{-\frac c2-s+1}.
\ee
Moreover we will expand the ratio of zeta functions
\be
\frac{\zeta(2s)}{\zeta(2s-1)} = \sum_{n=1}^{\infty} b(n) n^{-2s},
\ee
where $b(n)$ is a number-theoretic function defined as
\be
b(n) \coloneqq \sum_{k | n} k\mu(k).
\ee 
where $\mu(n)$ is the M\"obius function:
\be
\mu(n) \coloneqq \begin{cases} (-1)^{\text{number of prime factors of $n$}} ~~~~ &\text{$n$ is square-free} \\ 0 & \text{$n$ is divisible by a prime squared.} \end{cases}
\ee
We can then rewrite (\ref{eq:gammastuff}) as
\begin{align}
\int_{\gamma-i\infty}^{\gamma+i\infty} &ds \pi^{s-\frac c2} \Gamma\left(\frac c2-s\right)\mathcal{E}_{\frac c2-s}^c(\mu)y^{s-\frac c2} \nonumber \\ &= \sum_{\Delta \in \mathcal{S}} \sum_{n=1}^{\infty} b(n) \int_{\gamma-i\infty}^{\gamma+i\infty} ds \frac{\Gamma(s)\Gamma(s+\frac c2-1)}{\pi^{s+\frac{c-1}2}\Gamma(s-\frac 12)} (2\Delta)^{-\frac c2-s+1} y^{s-\frac c2} n^{-2s}.
\label{eq:thisgivesU}
\end{align}
The integral in (\ref{eq:thisgivesU}) is related to a confluent hypergeometric function of the second kind %(whatever that means)
(see 13.4.18 of \cite{NIST:DLMF}), which we denote as $U$ (and is given by \texttt{HypergeometricU} in Mathematica):
\be
\frac1{2\pi i}\int_{\gamma-i\infty}^{\gamma+i\infty} ds \frac{\Gamma(s)\Gamma(s+\frac c2-1)}{\pi^{s+\frac{c-1}2}\Gamma(s-\frac 12)} (2\Delta)^{-\frac c2-s+1} y^{s-\frac c2} n^{-2s} = \frac{y^{1-c}}{\sqrt \pi} n^{c-2} U\left(-\frac12, \frac c2, \frac{2\pi n^2\Delta}y\right) e^{-\frac{2\pi n^2 \Delta}y}. 
\label{eq:thisIsU}
\ee
Thus we get a final crossing equation of:
\begin{empheq}[box=\fbox]{equation} \label{eq:crossingfinal}
\begin{split}
1 + \sum_{\Delta \in \mathcal{S}} e^{-2\pi \Delta y} &=  \frac{\Lambda\left(\frac{c-1}2\right)}{\Lambda\left(\frac c2\right)} y^{1-c} + \varepsilon_c(\mu) y^{-\frac{c}2} + \sum_{k=1}^\infty \text{Re}\left(\delta_{k,c}(\mu)y^{-\frac c2 + 1 - \frac{z_k}2}\right) \\
&+ \frac{y^{1-c}}{\sqrt \pi} \sum_{\Delta \in \mathcal S} \sum_{n=1}^{\infty} b(n) n^{c-2} U\left(-\frac12, \frac c2, \frac{2\pi n^2\Delta}y\right) e^{-\frac{2\pi n^2 \Delta}y}.
\end{split}
\end{empheq}
In addition to a rigorous derivation we have also numerically checked (\ref{eq:crossingfinal}) for various values of $c, y$ to a precision of $1$ part in $10^{70}$.

Another consistency check of (\ref{eq:crossingfinal}) one can perform analytically is to consider the large $y$ limit. In this limit, the LHS is dominated by $1$ from the identity, but each term on the RHS is perturbatively small at large $y$. Similar to the lightcone bootstrap of four-point functions \cite{Fitzpatrick:2012yx,Komargodski:2012ek}, it turns out that the leading term on the LHS is reproduced by the infinite sum over $\Delta$ in the RHS. More precisely, one can show that
\be
\frac{y^{1-c}}{\sqrt{\pi}}\sum_{n=0}^{\infty}b(n)n^{c-2}\int_0^{\infty} d\Delta \frac{2\pi^c\zeta(c-1)\Delta^{c-2}}{\zeta(c)\Gamma(\tfrac c2)^2}U\left(-\frac12, \frac c2, \frac{2\pi n^2\Delta}y\right) e^{-\frac{2\pi n^2 \Delta}y} =1
\ee
where $\frac{2\pi^c\zeta(c-1)\Delta^{c-2}}{\zeta(c)\Gamma(\tfrac c2)^2}$ is the leading large $\Delta$ behavior of the spectral density (and which is the average spectral density for Narain theories; see \cite{Afkhami-Jeddi:2020ezh, Maloney:2020nni}). It might also be interesting to understand how the perturbatively small terms at large $y$ on the RHS of (\ref{eq:crossingfinal}) cancel among each other to give the non-perturbatively small corrections on the LHS.

\subsection{Functionals}
\label{sec:functionalbohnanza}

We would now like to apply linear functionals to (\ref{eq:crossingfinal}) to obtain sum rules that can constrain the possible sets $\mathcal{S}$. In particular we would like to put a bound on the scalar gap, meaning the lightest operator present in $\mathcal{S}$. One immediate problem is that not every term in (\ref{eq:crossingfinal}) is sign-definite. The term $\varepsilon_c(\mu)$ is not sign-definite, and the infinite terms $\delta_{k,c}(\mu)$ are also not sign-definite for any $k$. To remove the $\varepsilon_c(\mu)$ term is straightforward. Let us start by rewriting (\ref{eq:crossingfinal}) as:
\begin{align}
    \sum_{\Delta \in \mathcal{S}}&\left[y^{\frac c2}e^{-2\pi \Delta y} - \frac{y^{1-\frac c2}}{\sqrt \pi} \sum_{n=1}^{\infty} b(n) n^{c-2} U\left(-\frac12, \frac c2, \frac{2\pi n^2 \Delta}y\right) e^{-\frac{2\pi n^2 \Delta}y}\right] \nn\\
    &= - y^{\frac{c}{2}}+\frac{\Lambda\left(\frac{c-1}2\right)}{\Lambda\left(\frac c2\right)} y^{1-\frac c2} + \varepsilon_c(\mu) + \sum_{k=1}^\infty \text{Re}\left(\delta_{k,c}(\mu) y^{1-\frac{z_k}2}\right).
    \label{eq:ca}
\end{align} 
Taking a derivative with respect to $y$ removes the $\varepsilon_c(\mu)$ term. If we then redefine $t^2 \coloneqq y^{-1}$ we get:
\begin{align}
    \sum_{\Delta \in \mathcal{S}} &\Bigg[t^{-c}(4\pi \Delta - c t^2) e^{-\frac{2\pi \Delta}{t^2}}- \frac{t^c}{\sqrt \pi} \sum_{n=1}^\infty b(n) n^{c-2} e^{-2\pi \Delta n^2 t^2}\times \nn\\&
    \left((c-2-4\pi n^2 t^2 \Delta) U\left(-\frac12, \frac c2, 2\pi n^2 t^2 \Delta\right) + 2\pi n^2 \Delta t^2 U\left(\frac12, \frac c2+1, 2\pi n^2 t^2 \Delta\right)\right)\Bigg] \nn\\
    &=  c t^{2-c} + \frac{\Lambda\left(\frac{c-1}2\right)}{\Lambda\left(\frac c2\right)} (c-2) t^c + \sum_{k=1}^\infty \text{Re}\left(\delta_{k,c}(\mu) (z_k-2)t^{z_k}\right).
\label{eq:cb}
\end{align}

Now we need a functional acting on (\ref{eq:cb}) to remove terms of the form $t^{z_k}$ where $z_k$ is a nontrivial zero of the Riemann zeta function. To accomplish this we use the following family of functionals\footnote{We are extremely grateful to Danylo Radchenko for explaining this strategy to us. See \cite{Radchenko} for further generalizations of this. The construction of the functionals in \cite{Radchenko} seems to be reminiscent of the analytic functionals in \cite{Mazac:2016qev}. It might be interesting to explore the connection further.}.

Consider an even function $\varphi(t)$ that satisfies the following properties:
\begin{itemize}
    \item $\varphi(t)$ and $\hat \varphi(t)$ both decay rapidly (faster than any polynomial) at infinity
    \item $\varphi(t)$ and $\hat \varphi(t)$ have no singularities at finite $t$
    \item $\varphi(0) = \hat\varphi(0) = 0$
    \item $\int_0^{\infty}\frac{dt}{t}\varphi(t)t^s$ admits an analytic continuation to all $s\in \mathbb{C}$ (which we will call $M_{\varphi}(s)$),
\end{itemize}
where $\hat\varphi$ is the Fourier transform of $\varphi$:
\be
\hat\varphi(p) \coloneqq \int_{-\infty}^{\infty} dx~ e^{-2\pi i p x} \varphi(x).
\ee
We define
\be
\Phi(t) \coloneqq \sum_{n=1}^\infty \varphi(n t).
\label{eq:Phideff}
\ee
The function $\Phi(t)$ can also be rewritten via the Poisson resummation formula as
\begin{align}
\Phi(t) &= -\frac12 \varphi(0) + \frac{1}{2t} \hat\varphi(0) + \frac 1t \sum_{n=1}^\infty \hat\varphi\left(\frac nt\right) \nn\\
&= \frac 1t \sum_{n=1}^{\infty} \hat\varphi\left(\frac nt\right).
\label{eq:Phi_Poisson}
\end{align}
Combining (\ref{eq:Phideff}) and (\ref{eq:Phi_Poisson}) and the properties listed above, we see that $\Phi(t)$ decays faster than any polynomial at both small $t$ and large $t$.

Now, we define a functional $\mathcal{F}^\varphi[h(t)]$ by
\be
\mathcal{F}^\varphi[h(t)] \coloneqq \int_0^\infty \frac{dt}t h(t) \Phi(t).
\label{eq:deffunc}
\ee
Let us first consider the action of the functional on a power of $t$:
\be
\mathcal{F}^\varphi[t^s] = \int_0^\infty \frac{dt}t t^s \Phi(t).
\ee
Because of the properties of $\Phi(t)$ discussed above, $\mathcal{F}^\varphi[t^s]$ is an analytic function on the entire complex $s$ plane. Moreover, for $\Re (s) >1$, we can exchange the integration and the summation, which gives
\begin{align}
    \mathcal{F}^\varphi[t^{s}] &= \int_0^\infty dt ~t^{s-1} \Phi(t) \nn\\ &= \int_0^\infty dt ~t^{s-1} \sum_{n=1}^\infty \varphi(nt) \nn\\
    &= \sum_{n=1}^\infty n^{-s} \int_0^\infty dt ~t^{s-1} \varphi(t) \nn\\
    &= \zeta(s) \int_0^\infty dt~ t^{s-1} \varphi(t),\qquad \qquad \Re (s)>1.
    \label{eq:zetavanish0}
\end{align}
Properties of analytic continuation then imply that for all $s\in \mathbb{C}$,
\be
\mathcal{F}^\varphi[t^s] = \zeta(s)M_{\varphi}(s).
\label{eq:zetavanish}
\ee

From (\ref{eq:zetavanish}) we see that the functional $\mathcal{F}^\varphi$ will remove the final sign-indefinite terms $\delta_{k,c} t^{z_k}$ in our crossing equation (\ref{eq:cb}). We can then apply the functional $\mathcal{F}^\varphi$ to (\ref{eq:cb}) to get a positive sum rule the scalar operators must satisfy. Let us consider the situation where $\varphi(t)$ is a (finite) linear combination of Gaussians, for which $M_{\varphi}(s)$ is a sum of Gamma functions. In particular we consider the following family of $\varphi(t)$ defining the functionals:
\begin{align}
    \varphi(t) &= \sum_{i=1}^N \alpha_i e^{-\pi k_i t^2},
    \label{eq:varphidef}
\end{align}
where $k_i, \alpha_i$ are an arbitrary set of $N$ real numbers. In order for $\varphi(t)$ to satisfy $\varphi(0)=\hat \varphi(0)=0$, we choose $k_i, \alpha_i$ subject to the constraints
\begin{align}
\sum_{i=1}^N \alpha_i &= 0, \nn\\
\sum_{i=1}^N \alpha_i k_i^{-1/2} &= 0.
\label{eq:sumconstraint}
\end{align}
With this definition of $\varphi$, we can define $\Phi$ and the action of the functional $\mathcal{F}$ by using (\ref{eq:Phideff}) and (\ref{eq:deffunc}). If we then apply this functional to our crossing equation (\ref{eq:cb}), we get a positive sum rule for the operators $\Delta$. In Appendix \ref{sec:morefunc}, we write down explicit formulas for the action of this functional on (\ref{eq:cb}) with a single Gaussian $\varphi(t) = e^{-\pi k t^2}$ as a function of $\Delta$ and $k$.

Although in principle we could choose any functional via (\ref{eq:varphidef}) obeying (\ref{eq:sumconstraint}), for numerical calculations it will be more convenient to use functionals consisting of derivatives with respect to $k$, evaluated at $k=1$ instead. To be more explicit, the sum rule we get after applying the functional from (\ref{eq:varphidef}) is given by
\be
\sum_{i=1}^{N}\alpha_i \text{vac}(k_i) + \sum_{i=1}^{N}\sum_{\Delta}\alpha_i f(k_i, \Delta) =0,
\label{eq:generalsumrule}
\ee
subject to the constraints (\ref{eq:sumconstraint}). $f(k,\Delta)$ and $-\text{vac}(k)$ are the actions of the functional on the LHS and RHS respectively of (\ref{eq:cb}) (with explicit formulas given in Appendix \ref{sec:morefunc}, see e.g. (\ref{eq:cb3})). Let us consider the action of a single Gaussian of width $k$ (i.e not yet obeying the constraints above):
\be
\text{vac}(k) + \sum_{\Delta} f(k, \Delta).
\label{eq:2good2btrue}
\ee
The expression (\ref{eq:2good2btrue}) is not equal to $0$ because we have not obeyed the constraints (\ref{eq:sumconstraint}). However, the only functions of $k$ that it can be equal to are a constant term and a term proportional to $k^{-1/2}$. Any other term would allow some combination of functionals obeying (\ref{eq:sumconstraint}) to not vanish, and thus contradict (\ref{eq:generalsumrule}). Therefore we have
\be
\text{vac}(k) + \sum_{\Delta} f(k, \Delta) = c_0 + c_1 k^{-1/2},
\label{eq:2good2bfalse}
\ee
where $c_0, c_1$ are $k$-independent constants (they could be theory-dependent however). From an explicit calculation of $\text{vac}(k)$ and $f(k,\Delta)$ in Appendix \ref{sec:morefunc}, we see that
\be
\text{vac}'(1) = \partial_k f(k,\Delta)|_{k=1} = 0, 
\label{eq:thisguyvanishes}
\ee
which implies $c_1=0$.\footnote{In fact it turns out that $c_0$ is related to $\varepsilon_c(\mu)$ (defined in (\ref{eq:residuedelta})) via $c_0 = \frac{\pi \varepsilon_c(\mu)}6$. This in principle leads to a stronger crossing equation but we find that numerically it gives very similar bounds on the scalar gap, so we will not explore it further in this paper.} Thus we have
\be \text{vac}^{(n)}(1) + \sum_{\Delta} (\partial_k)^n f(k, \Delta)|_{k=1} = 0, ~~n\geq 2 
\label{eq:basistouse}
\ee
which will be the basis for our functionals. (Only even values of $n$ will provide independent equations, however.)

Notice that
\be
\text{vac}^{(n)}(1) = \lim_{\Delta\rightarrow0} (\partial_k)^n f(k,\Delta)|_{k=1},~~ n \geq 2
\ee
so indeed the vac term in (\ref{eq:basistouse}) is precisely the contribution of the vacuum ($\Delta=0$) to the sum rule (and the same is true in (\ref{eq:generalsumrule})).

\subsection{Numerical results}
\label{sec:numerics}

In this section, we present the numerical results for bounds on the scalar gap of $U(1)^c$ CFTs for various values of $c$ obtained from using the basis of functionals (\ref{eq:basistouse}). Note that the hypergeometric function in (\ref{eq:crossingfinal}) for odd values of $c$ reduces to an elementary function, which greatly simplifies the technical calculations. We therefore focus on odd values of $c$ (although there is nothing in principle stopping the following from working for even $c$). We first consider the functional obtained from taking $2$ and $4$ derivatives of (\ref{eq:basistouse}), and obtain a bound on $\Delta_{\text{scalar gap}}$ from these two sum rules, following the approach in \cite{Hellerman:2009bu}. Since we take at most 4 derivatives, we denote this bound as $\Delta^{(4)}_{\text{scalar gap}}$ (and more generally define a bound from at most $n$ derivatives as $\Delta^{(n)}_{\text{scalar gap}}$). Note that $\Delta^{(n)}_{\text{scalar gap}}$ is obtained from $\frac n2$ functionals.

\begin{figure}[t]
\begin{center}
  \includegraphics[width=0.9\linewidth]{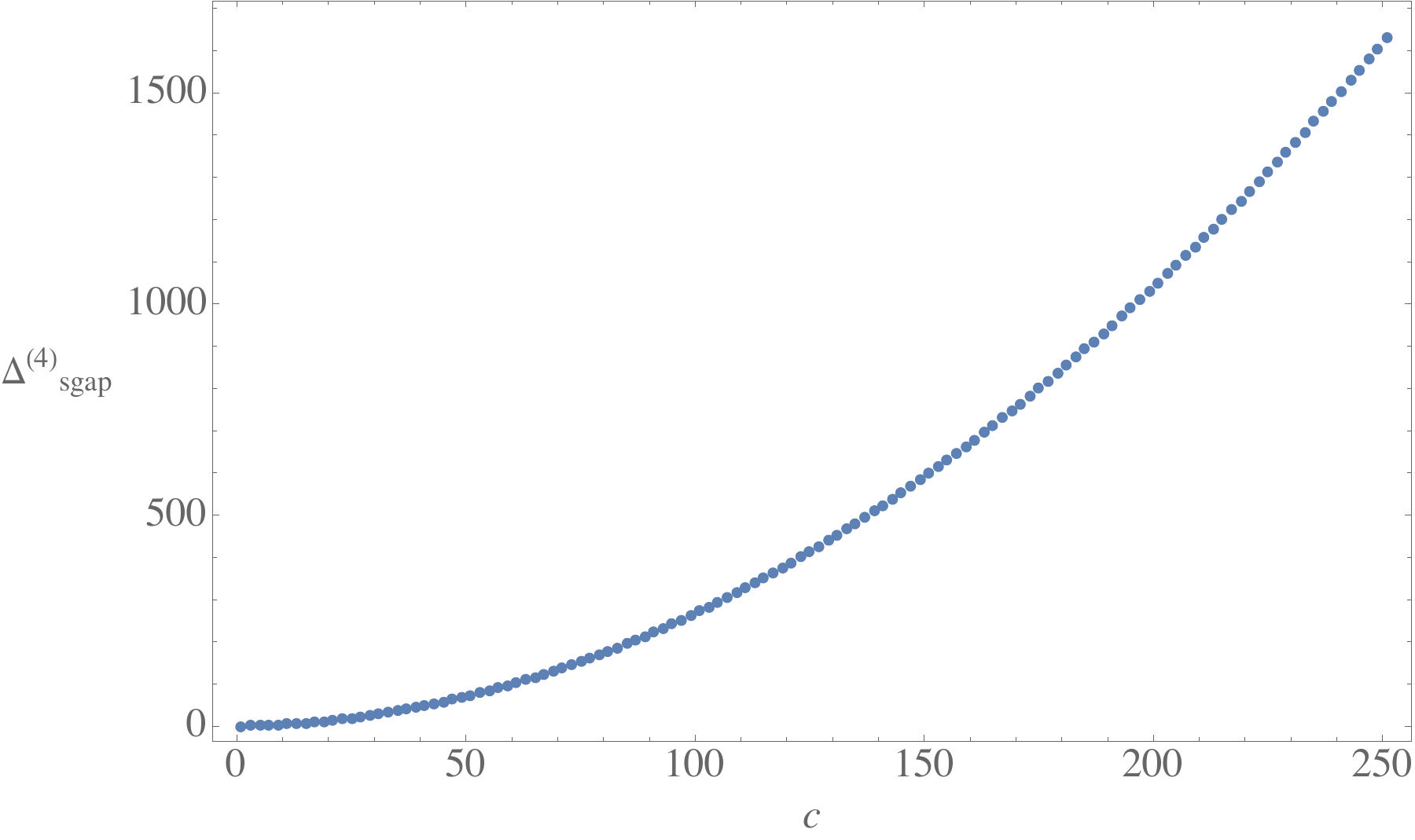}
  \end{center}
  \caption{Plot of a bound on the scalar gap for $U(1)^c$ CFTs with 4 derivatives, up to central charge $c=251$. The numerical data seems to be well-approximated by a quadratic function with leading coefficient $\frac1{4\pi^2}$ (see (\ref{eq:fitthedata})).}
  \label{fig:plotu1twoderiv}
\end{figure}

We have computed  $\Delta^{(4)}_{\text{scalar gap}}$ for odd central charge up to $251$.\footnote{At $c=1$ the crossing equation we use is slightly different due to a divergence of the zeta function at $1$; see Appendix \ref{sec:crossingc1subtlety} for discussion.} The results are plotted in Fig. \ref{fig:plotu1twoderiv}. The bound at large $c$ numerically appears to grow quadratically with $c$. Fitting it to a quadratic function gives
\be
\Delta^{(4)}_{\text{scalar gap}}(c) \sim 0.0253303c^2+0.13506c+0.400.
\label{eq:fitthedata}
\ee
The coefficient of the leading term is very close to $\frac{1}{4\pi^2}\approx 0.0253302959$. It may be possible to analytically prove that $\Delta^{(4)}_{\text{scalar gap}}(c) \sim \frac{c^2}{4\pi^2}$ at large $c$. Note that in this analysis we only considered 4 derivatives of (\ref{eq:basistouse}), but it may be the case that if we take $c \rightarrow \infty$ with fixed number of derivatives, the leading asymptotics for the bound is independent of the number of derivatives. This is indeed what happens in the spinless modular bootstrap, where the large $c$ bound at any fixed number of derivatives scales as $\frac c6$ \cite{Friedan:2013cba}.

It would be better to do the analysis with the opposite order of limits, where we take the number of derivatives to large before taking $c$ large (as in \cite{Collier:2016cls}) and then extrapolate in $c$. We can obtain bounds from including a larger number of derivatives in (\ref{eq:basistouse}) by using the semidefinite program solver SDPB \cite{Simmons-Duffin:2015qma,Landry:2019qug}. More precisely, we consider the sum rule
\be
\sum_{n=2,4,\ldots,n_{\mathrm{max}}} \alpha_n \text{vac}^{(n)}(1) + \sum_{n=2,4,\ldots,n_{\mathrm{max}}} \alpha_n \sum_{\Delta}(\partial_k)^{n}f(k,\Delta)|_{k=1} =0.
\ee
Unfortunately, the function $(\partial_k)^{n}f(k,\Delta)|_{k=1}$ in (\ref{eq:basistouse}) does not have a good approximation as a product of a positive function of $\Delta$ and a polynomial in $\Delta$. Therefore, we discretize in $\Delta$-space and sample the function $(\partial_k)^{n}f(k,\Delta)|_{k=1}$ at various points $\Delta_1,\Delta_2,\ldots,\Delta_{M}$ above the scalar gap assumption, and use SDPB as a linear programming solver to look for a functional that satisfies
\begin{align}
&\sum_{n=2,4,\ldots,n_{\mathrm{max}}} \alpha_n \text{vac}^{(n)}(1) =1, \nonumber \\
&\sum_{n=2,4,\ldots,n_{\mathrm{max}}} \alpha_n(\partial_k)^{n}f(k,\Delta)|_{k=1} \geq 0,\qquad \Delta=\Delta_1,\ldots,\Delta_M.
\end{align}
Finally, we check the positivity of the obtained functional for all $\Delta \geq \Delta^{(n_{\mathrm{max}})}_{\text{scalar gap}}$ by hand. If there is a negative region, we sample more points there and rerun SDPB, and repeat this procedure until the functional is positive or SDPB gives a primal feasible solution\footnote{We are extremely grateful to David Simmons-Duffin for explaining this approach to us.}.

\begin{table}[t]
\begin{center}
    \begin{tabular}{| c | c | c | c| c| c | c || c | }
    \hline
    $c$ &$\Delta_{\text{scalar gap}}^{(10)}$&$\Delta_{\text{scalar gap}}^{(20)}$ &$\Delta_{\text{scalar gap}}^{(30)}$ &$\Delta_{\text{scalar gap}}^{(40)}$ & $\Delta_{\text{scalar gap}}^{(50)}$ & $\Delta_{\text{scalar gap}}^{(60)}$ & $\Delta_{\text{avg sgap}}$  \\ \hline
    1 & 0.507 & $\frac {1}{2} + 7 \times 10^{-5}$ & $\frac12 + 2 \times 10^{-6} $& $\approx \frac12$ & $\approx \frac12$ & $\approx \frac12$ & ill-defined \\
    3 & 0.910 & 0.864 & 0.863 & 0.863 & 0.863 & 0.863 & 0.136  \\
    5 & 1.444 & 1.310 & 1.304 & 1.303 & 1.302 & 1.302 & 0.324 \\
    7 & 2.129 & 1.843 & 1.820 & 1.814 & 1.813 & 1.813 & 0.471 \\ 
    9 & 2.972 & 2.476 & 2.419 & 2.400 & 2.397 & 2.396 & 0.606 \\ 
    11 & 3.980 & 3.219 & 3.110 & 3.063 & 3.055 & 3.051 & 0.736 \\
    13 & 5.155 & 4.078 & 3.897 & 3.808 & 3.789 & 3.779 & 0.863\\
    15 & 6.500 & 5.058 & 4.788 & 4.638 & 4.602 & 4.581 & 0.989 \\
    17 & 8.018 & 6.614 & 5.786 & 5.558 & 5.497 & 5.458 & 1.113 \\
    19 & 9.709 & 7.399 & 6.895 & 6.570 & 6.477 & 6.412 & 1.237 \\
    21 & 11.576 & 8.765 & 8.118 & 7.680 & 7.545 & 7.445 & 1.360\\
    23 & 13.619 & 10.266 & 9.460 & 8.890 & 8.705 & 8.561 & 1.482 \\ 
    25 & 15.839 & 11.903 & 10.922 & 10.202 & 9.959 & 9.762 & 1.604 \\
    27 & 18.238 & 13.679 & 12.506 & 11.620 & 11.310 & 11.049 & 1.725 \\ \hline
    \end{tabular}
    \caption{Upper bounds on the scalar gap from $U(1)^c$ CFTs with odd $c \leq 27$ after taking up to $10, 20, \cdots, 60$ derivatives of our crossing equation (i.e. the maximum value of $n$ in (\ref{eq:basistouse})) computed to three decimal places. We also compare it to the average Narain scalar gap, defined in (\ref{eq:avg}) (though note that the optimal bound is different from the average). See Fig. \ref{fig:plotu1} for a plot.}
    \label{tab:table}
\end{center} 
\end{table}

\begin{figure}[t]
\begin{center}
  \includegraphics[width=1.0\linewidth]{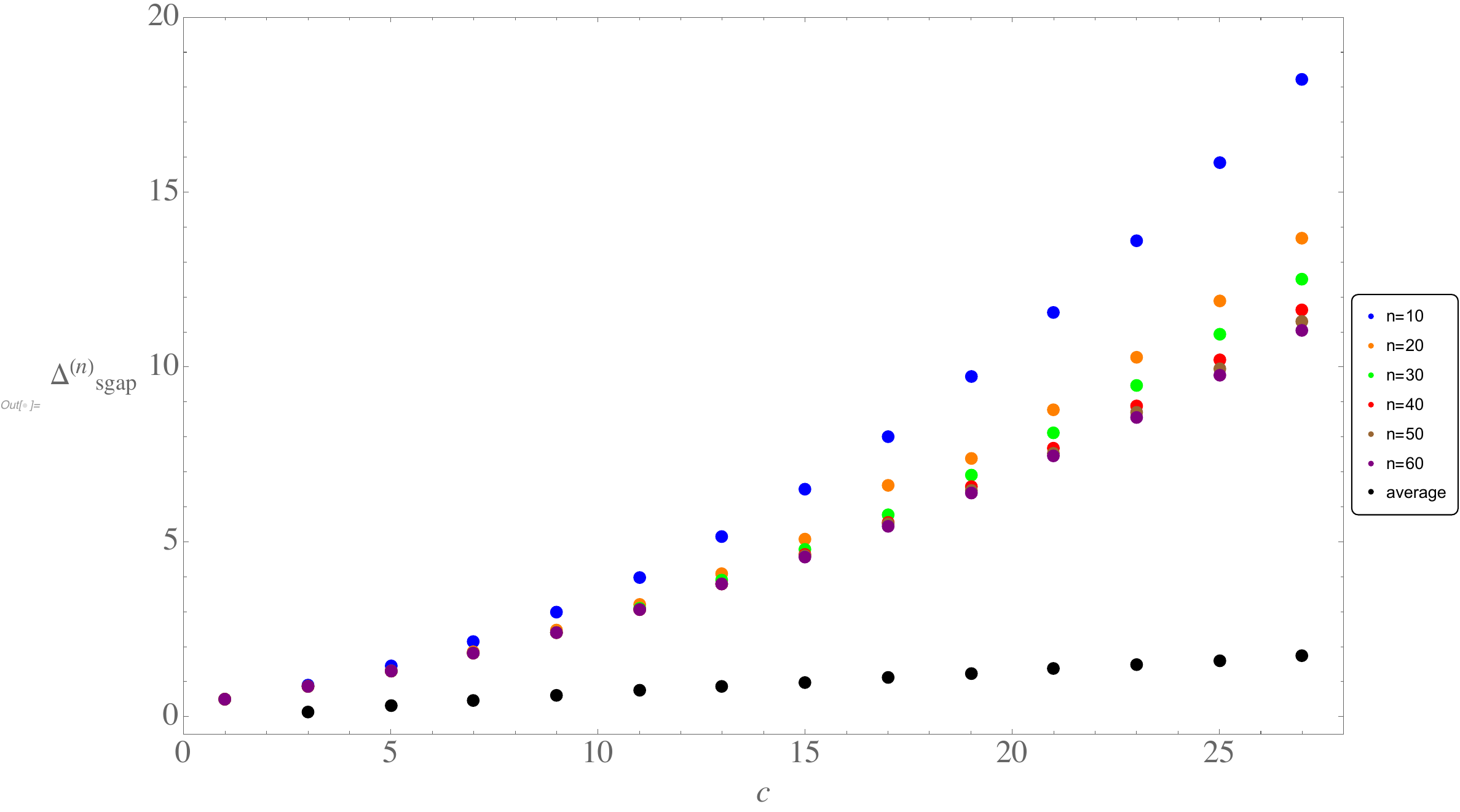}
  \end{center}
  \caption{Plot of a bound on the scalar gap for $U(1)^c$ CFTs at odd $c \leq 27$. The colors blue, orange, green, red, brown, and purple represent the bound we get at $10, 20, \cdots, 60$ derivatives respectively. The color black represents the average Narain scalar gap, for comparison. (However, there is no a priori reason the average Narain scalar gap and the optimal $U(1)^c$ scalar gap should be similar.) See Table \ref{tab:table} for the numerical data.}
  \label{fig:plotu1}
\end{figure}

Using the method described above, we have computed $\Delta^{(n)}_{\text{scalar gap}}$ for $n=10, 20, \cdots, 60$ for central charge odd $c \leq 27$. Our bounds are summarized in Table \ref{tab:table} and plotted in Fig. \ref{fig:plotu1}.\footnote{In Table 1 of \cite{Afkhami-Jeddi:2020ezh}, a bound on the gap (of any spin) was computed using the spinless modular bootstrap. Our results in Table \ref{tab:table} are specifically for scalars, and so in general are orthogonal. However, for $c=3$, the bound in \cite{Afkhami-Jeddi:2020ezh} is less than $1$ and so must be a scalar, and is stronger than the bound we found at $c=3$.} We were not able to go to high enough central charge to do a reliable extrapolation to large $c$. There are two obstacles in going to large central charge. The first is that the convergence of the bound as the derivative order $n\to \infty$ becomes slower for larger $c$. The second obstacle is that the number of terms in the sum rule (\ref{eq:basistouse}) grows as $c^4$ (see the sum in (\ref{eq:yay})), which makes evaluating derivatives with respect to $k$ very slow. It would be good if there were a more efficient way to compute the derivatives.

It is interesting to compare the bounds on the $U(1)^c$ scalar gap we get to the average Narain scalar gap. In \cite{Afkhami-Jeddi:2020ezh} an expression for the average scalar gap of Narain theories was computed, by first calculating the average density of states for all Narain theories (under the Zamolodchikov measure), and determining when the integral of the average density of states is $1$. By looking at the average density of scalars, \cite{Afkhami-Jeddi:2020ezh} got an average scalar gap of \footnote{Note that choosing the integrated average to be $1$, as opposed to any other $O(1)$ number less than $1$, is somewhat of a convention. However, if we choose another cutoff, the result (\ref{eq:avg}) changes very little.}
\begin{align}
\Delta_{\text{avg sgap}} &= \left(\frac{\zeta(c)\Gamma\left(\frac c2\right)^2(c-1)}{\zeta(c-1)2\pi^c}\right)^{\frac1{c-1}} \nonumber \\
&= \frac{c}{2\pi e} + \frac{\log c}{2\pi e} + O(1).
\label{eq:avg}
\end{align}
Our numerical bounds at large $c$ (including our bounds with four derivatives extrapolated to large $c$) appear to be very far from both the average Narain scalar gap and the bound on the gap of the lightest operator of any spin (see (\ref{eq:u1gap})). It would be interesting to explore further if our bounds on the scalar gap can be substantially improved by considering other crossing equations. Of course, it is possible that the optimal scalar gap behaves differently from both the average Narain scalar gap and the optimal gap at large $c$.

\section{General 2d CFTs}
\label{sec:virasoro}

So far our discussion has been restricted to a very special class of CFTs, namely those with $U(1)^c$ chiral algebra. In this section we generalize to generic 2d CFTs, which only have Virasoro symmetry and no extended chiral algebra (though we pause to note that we do not have any explicit examples of such theories, even numerically \cite{Collier:2016cls}).

The main obstacle to repeating our analysis to general 2d CFTs is that the partition function is not square-integrable, due to the Casimir energy of the theory on a cylinder. For theories with $U(1)^c$ chiral algebra, when we factored out the characters of the theory and considered the primary counting partition function $\hat{Z}$, the resulting function grew only polynomially ($\sim y^{c/2}$) at the cusp (see (\ref{eq:primarycounting})). For theories with only Virasoro symmetry, however, the (Virasoro) primary counting partition function will grow as $\sim e^{2\pi \frac{c-1}{12} y}$ at large $y$. Although there are various ways we can get around this (see Sec. 4 of \cite{Benjamin:2021ygh} for some discussions of other approaches), in this section we will simply take the partition function multiply by the same cusp form as we did for theories with $U(1)^c$ symmetry, and bound the resulting function we get. This will not give us a crossing equation acting only on the Virasoro scalar operators, but instead will give us an equation acting on a more complicated combination of operators of all spin. 

To be more precise, let us consider any compact 2d CFT with $c>1$ and only Virasoro symmetry as its maximal chiral algebra (although generalizations to other chiral algebras are simple). Suppose the partition function of this theory is $Z(\tau)$. We define the ``fake scalars" of this theory as
\be
Z^{\text{fake scalars}}(y) = \int_{-1/2}^{1/2} dx |\eta(\tau)|^{2c} Z(\tau). 
\label{eq:fakescalars}
\ee
Note that the central charge $c$ is not necessarily an integer in this analysis. We call this function ``fake scalars" because if this theory were to have a $U(1)^c$ chiral algebra, then (\ref{eq:fakescalars}) would be a count of the scalars (under the $U(1)^c$ algebra). However, since the theory only has Virasoro symmetry, then $Z^{\text{fake scalars}}(y)$ does not in general have a positive $q$-expansion.

Even without the full $U(1)^c$ chiral algebra, the logic in deriving the crossing equation (\ref{eq:crossingfinal}) in Sec. \ref{sec:narain} will apply to $Z^{\text{fake scalars}}(y)$. We can still apply harmonic analysis to $y^{c/2} |\eta(\tau)|^{2c} Z(\tau) - E_{c/2}(\tau)$ and derive an analogous crossing equation for $Z^{\text{fake scalars}}(y)$. To be precise, the equation we derive is the following.

Let $a_c(n)$ be defined as\footnote{At central charge $25$, $a_{c=25}(n)$ is the Ramanujan tau function (up to a shift of the argument by $1$).}
\be
\sum_{n=0}^{\infty} a_c(n) q^n = \prod_{n=1}^{\infty} (1-q^n)^{c-1}.
\ee
Then we have the following crossing equation in terms of the Virasoro primary operators of any $c>1$ compact CFT:
\begin{align}
&\sum_{n=0}^{\infty} e^{-4\pi y n} a_c(n)^2 + \sum_{\Delta, j \in \mathcal{S} \cup \mathcal{S}^{\text{null}}} \sum_{n=0}^{\infty} e^{-2\pi y (\Delta + j + 2n)} a_c(n) a_c(n+j) =\nn\\
&\frac{\Lambda\(\frac{c-1}2\)}{\Lambda\(\frac c2\)} y^{1-c} + \varepsilon y^{-\frac c2} + \sum_{n=1}^{\infty} \text{Re}\left[\delta_n y^{-\frac c2 + 1 - \frac{z_n}2}\right] \nn\\
&~~~+ \frac{y^{1-c}}{\sqrt \pi} \sum_{\Delta, j \in \mathcal{S} \cup \mathcal{S}^{\text{null}}}  \sum_{n=0}^{\infty} \sum_{k=1}^{\infty} b(k) k^{c-2} U\(-\frac12, \frac c2, \frac{2\pi k^2(\Delta + j + 2n)}{y}\) e^{-\frac{2\pi k^2(\Delta + j + 2n)}y} a_c(n) a_c(n+j) \nonumber \\
&~~~~+ \frac{y^{1-c}}{\sqrt{\pi}} \sum_{n=1}^{\infty} \sum_{k=1}^{\infty} b(k) k^{c-2} U\left(-\frac12, \frac c2, \frac{4\pi k^2 n}y\right) e^{-\frac{4\pi k^2 n}y} a_c(n)^2.
\label{eq:crossingggg}
\end{align}
In (\ref{eq:crossingggg}), $\mathcal{S}$ is the set of all non-vacuum Virasoro primary operators, labeled by their dimension $\Delta = h + \bar h$ and their spin $j = |h - \bar h|$. Moreover we define $\mathcal{S}^{\text{null}}$ formally as a set containing $-2$ operators of weight $1$, spin $1$ and $1$ operator of weight $2$, spin $0$. This is simply to take into account the level $1$ null state in the Virasoro vacuum block (i.e. that $L_{-1}$ and $\overline{L}_{-1}$ annihilate the vacuum). The LHS of (\ref{eq:crossingggg}) is precisely what we called $Z^{\text{fake scalars}}(y)$ above, written in terms of the Virasoro primary operators of the theory, which we denoted by the set $\mathcal{S}$. For convenience we have assumed the theory has no additional conserved currents, but it is simple to generalize (\ref{eq:crossingggg}) to allow for them. 

We have tested (\ref{eq:crossingggg}) numerically on the pure gravity partition function of \cite{Maloney:2007ud, Keller:2014xba}, which we will denote as $Z^{\text{MWK}}(\tau)$, at various values of the central charge. For simplicity we have ignored the null state at level $1$ (even though this leads to a inconsistent chiral algebra due to the lack of charged twist zero states \cite{Benjamin:2020swg}, the resulting function is still modular invariant with a gap to the first primary operator, so it will obey (\ref{eq:crossingggg}), without including the contribution from $\mathcal{S}^{\text{null}}$). Strictly speaking we glossed over a subtlety in deriving (\ref{eq:crossingggg}). When we derived (\ref{eq:crossingfinal}) we used the fact that $\mathcal{E}^c_s = \sum_{\Delta \in \mathcal{S}} \Delta^{-2s}$ for $\text{Re}(s) > c-1$ because the sum converges for those values of $s$. However if we define $\mathcal{E}^c_s$ analogously for the ``fake scalars," it could potentially be the case that there is no $s$ such that the sum converges, due to the Cardy growth of the Virasoro primary operators. Nonetheless, we numerically find that (\ref{eq:crossingggg}) is still satisfied. It might be interesting to present a more rigorous derivation of this step. 

We can then apply the same functionals on (\ref{eq:crossingggg}) as discussed in Sec. \ref{sec:functionalbohnanza} to remove the sign-indefinite terms related to the nontrivial zeros of the zeta function. This gives sum rules the CFT must satisfy, where now all operators (instead of just scalars) participate. For example, at $c=3$, and taking two derivatives in (\ref{eq:basistouse}), we get
\begin{align}
   & \sum_{\Delta, j \in \mathcal{S} \cup \mathcal{S}^{\text{null}}}\sum_{n=0}^\infty a_{c=3}(n) a_{c=3}(n+j) f(\Delta + j + 2n) + \sum_{n=1}^{\infty} a_{c=3}(n)^2 f(2n) = \frac \pi4,
   \label{eq:sumrulec3}
\end{align}
where 
\begin{align}
f(\Delta) \coloneqq \frac{\pi\left(-3+8\pi^2\Delta+(3+4\pi^2\Delta)\cosh(2\sqrt 2\pi \sqrt\Delta)-6\sqrt2\pi\sqrt\Delta\sinh(2\sqrt 2 \pi \sqrt\Delta)\right)}{8\sinh^4\left(\sqrt2 \pi \sqrt\Delta\right)}.
\label{eq:sumrulec3v2}
\end{align}
(This comes from evaluating $\partial_k^2|_{k=1}$ on (\ref{eq:c3sumrulewithks}).)
We can apply the same family of functionals discussed in Sec. \ref{sec:functionalbohnanza} to (\ref{eq:crossingggg}) more generally and try to derive bounds on the various quantities (e.g. scalar gap, gap, etc.) from this crossing equation. Unfortunately we run into two distinct issues that stop us from bounding generic theories.

First, we see that at large $\Delta$, (\ref{eq:sumrulec3v2}) falls off as $\sim e^{-2\pi\sqrt{2\Delta}}$. In fact from Appendix \ref{sec:morefunc} we see that regardless of the central charge or derivative order, the functionals used in Sec. \ref{sec:functionalbohnanza} fall off with the same leading asymptotics. However, the asymptotic growth of operators in $\mathcal{S}$ comes from the Cardy formula \cite{Cardy:1986ie} and is $\sim e^{2\pi\sqrt{\frac{\Delta(c-1)}3}}$. We thus see that if $c \geq 7$, the sum rule does not obviously converge. Note that for $U(1)^c$ CFTs this was not an issue because there, the asymptotic density of primary operators grew polynomially in $\Delta$ ($\sim \Delta^{c-2})$. As a check we have verified (\ref{eq:sumrulec3}) for $Z^{\text{MWK}}(\tau)$ at $c=3$, but the analogous computation at $c=9$ diverges (even though both obey (\ref{eq:crossingggg})). We have also verified (\ref{eq:sumrulec3}) for various rational CFTs with $c<7$ where we only decompose into Virasoro characters.

It is unfortunate that we only get a falloff in $e^{-\# \sqrt{\Delta}}$ in our sum rules. This only happened after we integrated against the function $\Phi(t)$ in (\ref{eq:deffunc}). Before this integral (e.g. in (\ref{eq:crossingfinal}) and (\ref{eq:crossingggg})), we had a falloff as $e^{-\# \Delta}$, which will always overwhelm the Cardy growth at any central charge. It would be interesting to see if there were another choice of functional that would both remove the sign-indefinite terms related to nontrivial zeros of the zeta function, but still preserve the faster falloff in dimension.

Second, the asymptotically large $\Delta$ behavior of 
\be
\sum_{n=0}^{\infty} a_{c}(n) a_{c}(n+j) f(\Delta + j +2n)
\ee
does not have fixed sign: for some spins the asymptotic $\Delta$ value is positive and for some spins it is negative. (This is true when one takes any number of derivatives of the crossing equation, not just two.) The root of this problem is that $a_c(n) a_c(n+j)$ does not have a definite sign. Thus there is no obvious way to construct functionals that have fixed sign for all spin and all dimensions larger than some cutoff.

We note that we have chosen to multiply the partition function by the cusp form $y^{c/2} |\eta(\tau)|^{2c}$ to render the partition function square-integrable. However any cusp form with a gap to the first excited state and that falls off at least as fast as $(q\bar{q})^{c/24}$ would be sufficient and give a similar crossing equation as (\ref{eq:crossingggg}). It might be useful to explore constraints one gets from other cusp forms.

Finally we end this section with an interesting observation. Our crossing equation (\ref{eq:crossingggg}) for Virasoro theories involves operators of all spins, since multiplying by $y^{c/2} |\eta(\tau)|^{2c}$ does not have an obvious physical interpretation for theories without a $U(1)^c$ extended current algebra. It would be better to have a crossing equation or sum rule that only involved scalar Virasoro primary operators. Surprisingly, we find strong hints that such a sum rule exists.

In order to get a sum rule acting only on scalar Virasoro primary operators, the natural thing to do is to multiply the partition function by $y^{1/2} |\eta(\tau)|^2$. This is the same object that we multiply for $U(1)^c$ theories for $c=1$. Recall that there, we derived the following sum rule (see (\ref{eq:betterCross}) and App. \ref{sec:crossingc1subtlety}):
\begin{align}
\log k + \sum_{\Delta\in\mathcal{S}} \left[ h(k, \Delta) - h(k^{-1}, \Delta) \right] = 0,
\label{eq:c1crossgeneralk}
\end{align}
where
\begin{align}
h(k,\Delta) \coloneqq \sqrt2 \pi \sqrt{k\Delta}(1-\text{coth}(\sqrt 2 \pi \sqrt{k\Delta}))+2\pi^2 k \Delta \text{csch}^2(\sqrt 2\pi \sqrt{k\Delta}) + \log(1-e^{-2\sqrt 2 \pi \sqrt{k \Delta}}).
\label{eq:ahkc1}
\end{align}
(The expression (\ref{eq:ahkc1}) is just (\ref{eq:hcgeneral}) at $c=1$, where we multiplied through by a factor of $-4$ for convenience.)

Remarkably, we numerically find that (\ref{eq:c1crossgeneralk}) also holds for general Virasoro CFTs, where $\mathcal{S}$ is now the set of conformal dimensions of scalar Virasoro primary operators (minus $\frac{c-1}{12}$) subject the following constraints. First of all, due to the null state structure of the Virasoro vacuum character, we introduce an additional term in $\mathcal{S}$ of $\Delta - \frac{c-1}{12} = -\frac{c-25}{12}$ (assuming no spin 1 currents). Second of all, we do not include the $\log k$ term in the sum rule (since there is not necessarily a state with $\Delta - \frac{c-1}{12} = 0$ in the spectrum). Finally, and most importantly, the sum rule does not converge for sufficiently large $c$. At large $\Delta$, we have
\be
h(k, \Delta) - h(k^{-1}, \Delta) \sim e^{-2\sqrt 2 \pi  \sqrt{\Delta \times \text{min}(k, k^{-1})}},
\ee
whereas
\be
\rho^{\text{scalar primaries}}(\Delta) \sim e^{2\pi \sqrt{\frac{\Delta(c-1)}3}},
\ee
so our sum rule only converges if
\be
c < 1 + 6 ~\text{min}(k, k^{-1}),
\label{eq:cconverge}
\ee
which implies $c < 7$.\footnote{Note also $h(k, \Delta) - h(k^{-1},\Delta)$ has poles at $\Delta = -\frac{n^2}{2k}$ and $\Delta = -\frac{n^2 k}2, n \in \mathbb N$, which may be problematic for convergence. For example, if $c=1+6 k n^2$ or $c=1 + 6k^{-1}n^2$, with $n \in \mathbb N$, then the vacuum term contributes as a pole.} 

For various theories obeying (\ref{eq:cconverge}), we very surprisingly find that the sum rule
\be
\int_{-\frac{c-1}{12}}^\infty d\Delta \rho^{\text{scalars}}(\Delta) \left[h(k, \Delta) - h(k^{-1}, \Delta)\right] = 0
\label{eq:sumrulecless7}
\ee
is obeyed to arbitrarily high precision. For $c<7$ we can use this to bound the Virasoro scalar gap. However, our bounds from this so far seem to be substantially weaker than those found in \cite{Collier:2016cls}. It would be extremely interesting if there were a way to analytically continue the sum in (\ref{eq:sumrulecless7}) to arbitrary central charge (and also to prove, or more honestly derive, (\ref{eq:sumrulecless7})). If so, this could be a way to derive a Virasoro scalar gap for all central charge\footnote{In \cite{Collier:2016cls}, it was shown that no bound on the Virasoro scalar gap could be derived for $c\geq25$ using the traditional modular bootstrap. This was due to the existence of a ``spurious solution" to crossing of $\frac{J(\tau) + \bar J(\bar\tau)}{\sqrt{\tau_2} |\eta(\tau)|^2}$, which lacks scalar primary operators (see discussion around Eqn (3.2) of \cite{Collier:2016cls}). However, if there exists a convergent sum rule like (\ref{eq:sumrulecless7}) for all $c$ that only acts on scalar primary operators, then by definition it would vanish on the spurious solution found in \cite{Collier:2016cls}, and one may be able to find a bound for $c\geq25$.}. 

\section{2d CFTs and the Riemann Hypothesis}
\label{sec:RH}

One interesting feature of our crossing equation (\ref{eq:crossingfinal}) is that in the small $y$ (high temperature) limit, the asymptotics are controlled by the real parts of the nontrivial zeros of the Riemann zeta function. Let us rewrite (\ref{eq:crossingfinal}), defining the temperature $T\coloneqq y^{-1}$, as
\begin{align}
1 + \sum_{\Delta \in \mathcal{S}} e^{-\frac{2\pi \Delta}T} &= \frac{\Lambda\left(\frac{c-1}{2}\right)}{\Lambda\left(\frac c2\right)} T^{c-1} + \varepsilon_c T^{\frac c2} \nn\\&+ \sum_{k=1}^\infty T^{\frac c2 - 1 + \frac{\text{Re}(z_k)}{2}} \left[\text{Re}(\delta_{k,c}) \cos(\text{Im}(z_k)\log T) - \text{Im}(\delta_{k,c})\sin(\text{Im}(z_k) \log T) \right] \nn\\&+ O\left(e^{-2\pi \Delta_{\text{gap}}T}\right).
\label{eq:RiemannStuff}
\end{align}
At high temperature, second line of (\ref{eq:RiemannStuff}) behaves as a highly oscillatory function with an overall envelope controlled by $\text{Re}(z_k)$. The Riemann hypothesis says that for all $k$,
\be
\text{Re}(z_k) = 1/2,
\ee
which would fix the envelope to be $T^{\frac c2 - \frac 34}$. In other words, if the Riemann hypothesis is true, (\ref{eq:RiemannStuff}) can be written as
\begin{align}
1 + \sum_{\Delta \in \mathcal{S}} e^{-\frac{2\pi \Delta}T} &= \frac{\Lambda\left(\frac{c-1}{2}\right)}{\Lambda\left(\frac c2\right)} T^{c-1} + \varepsilon_c T^{\frac c2} \nn\\&+ \sum_{k=1}^\infty T^{\frac c2 - \frac 34 } \left[\text{Re}(\delta_{k,c}) \cos(\text{Im}(z_k)\log T) - \text{Im}(\delta_{k,c})\sin(\text{Im}(z_k) \log T) \right] \nn\\&+ O\left(e^{-2\pi \Delta_{\text{gap}}T}\right).
\label{eq:RiemannStuff2}
\end{align}
However, if the Riemann hypothesis is false, then there is at least one $z_k$ with real part greater than $1/2$,\footnote{By the functional equation (\ref{eq:Lambdafunceq}) and meromorphicity, the Riemann hypothesis being false implies a pair of zeros of the zeta function with identical imaginary part: one with real part greater than $1/2$, one with real part less than $1/2$.} which changes the large temperature scaling in the second line of (\ref{eq:RiemannStuff2}).\footnote{Note that there is a possibility that the residue at that zero vanishes, meaning $\delta_{k,c}$ vanishes in (\ref{eq:RiemannStuff2}). However, this will only happen in a real codimension 2 subspace of the moduli space. Thus for a generic theory the scaling will change at large temperature. We thank Per Kraus for raising this question to us.\label{rhfoot}} Since the leading term of (\ref{eq:RiemannStuff}) is essentially the Cardy formula, then in some sense, the Riemann hypothesis makes a claim about the overall size of the ``subsubleading" corrections to the Cardy formula.

We can illustrate this with an explicit example. Let us consider the $SU(3)_1$ WZW model, and decompose the theory under the $U(1)^2$ chiral algebra (note that this is not the maximal chiral algebra). The scalar partition function is given by
\begin{align}
Z^{\text{scalars}}_{SU(3)_1}(T) &\coloneqq
1 + \sum_{\Delta \in \mathcal{S}_{SU(3)_1}} e^{-\frac{2\pi \Delta}T} \nn\\ 
&= 1 + \sum_{n=1}^\infty 48\left(\sum_{k|n} (-1)^k \sin(\tfrac{k \pi}{3})\right)^2 e^{-\tfrac{4\pi n}{T}}+ 24 \left(\sum_{k|3n-2} (-1)^k \sin(\tfrac{k \pi}{3})\right)^2 e^{-\tfrac{4\pi(n-\tfrac 23)}{T}}
\nn\\&= 1 + 18 e^{-\frac{4\pi}{3T}} + 36 e^{-\frac{4\pi} T} + 18 e^{-\frac{16\pi}{3T}} + 72 e^{-\frac{28\pi}{3T}} + \cdots.
\label{eq:su3scalars}
\end{align}
For $c=2$ the crossing equation (\ref{eq:RiemannStuff2}) is slightly modified due to a pole at $\Lambda(1/2)$. As derived in (\ref{eq:Exactcrossingc2}), the crossing equation we get for a $c=2$ Narain theory is
\begin{align}
    1 + \sum_{\Delta \in \mathcal{S}} e^{-\frac{2\pi \Delta}T} &=  \frac{3}{\pi} T \log T + \left[\hat{E_1}(\rho) + \hat{E_1}(\sigma) + \frac3\pi\(\gamma_E + \log(4\pi) + 24 \zeta'(-1) - 2\)\right]T  \nn\\
& + \sum_{k=1}^{\infty} \text{Re}\(\frac{4\pi^{\frac{z_k}{2}}  \Lambda(\frac{1+z_k}2)^2 E_{\frac{1+z_k}2}(\rho)  E_{\frac{1+z_k}2}(\sigma)}{2\Gamma\(\frac{z_k}2\)\zeta'(z_k)} T^{\frac{z_k}2}\)\nn\\
&+ \frac{T}{\sqrt\pi} \sum_{\Delta \in \mathcal{S}} \sum_{n=1}^\infty  b(n) U\(-\frac12, 1, 2\pi \Delta n^2 T\) e^{-2\pi \Delta n^2 T},
\label{eq:c2gen}
\end{align}
where $\hat {E_1}$ is defined in (\ref{eq:e1hat}). From the explicit form of the sum over $k$ in (\ref{eq:c2gen}), we see that the coefficient in front of $T^{\frac{z_k}2}$ falls off exponentially in $k$, so the sum converges rapidly.

\begin{figure}
\begin{center}
  \includegraphics[width=0.9\linewidth]{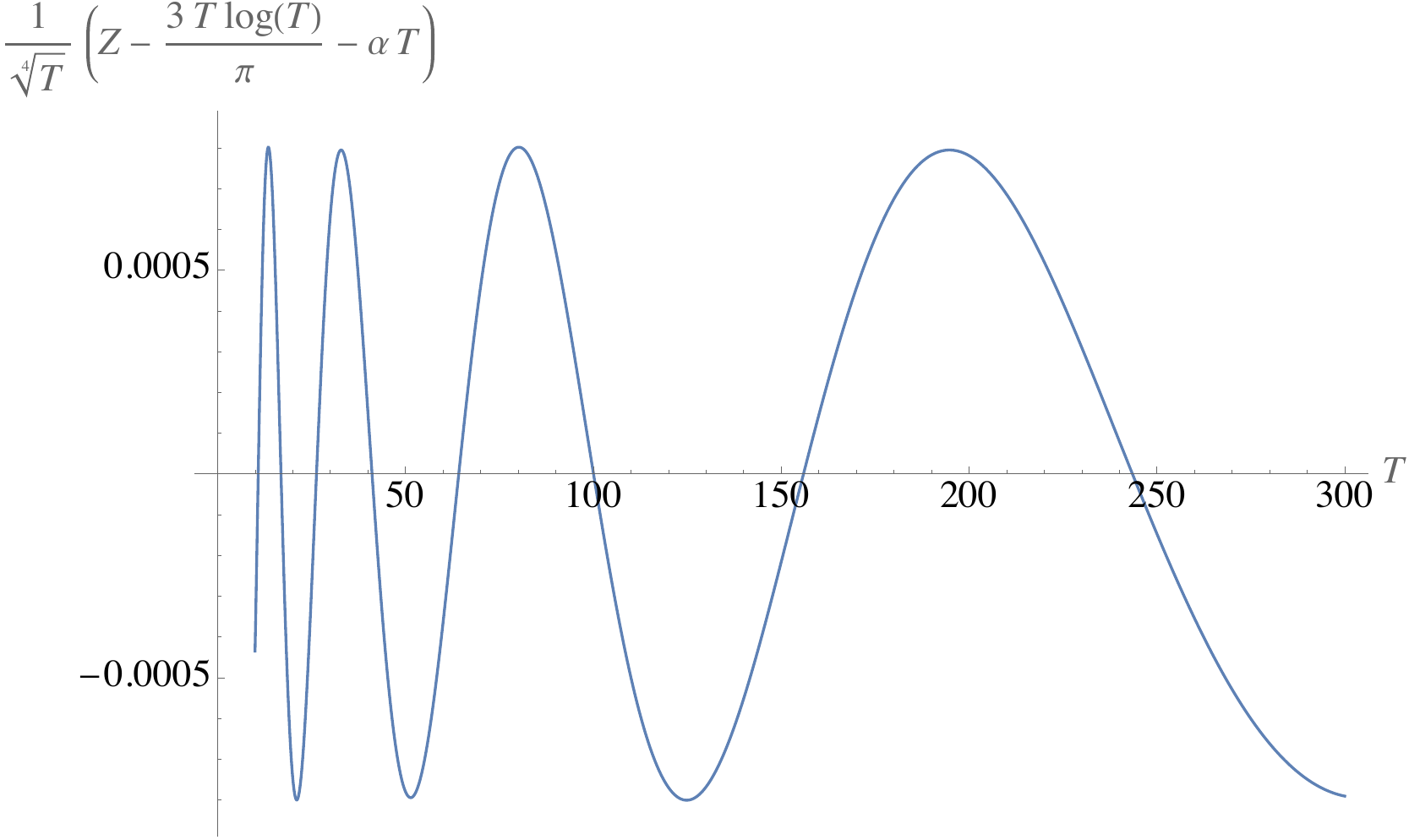}
  \end{center}
  \caption{Scalar part of the $SU(3)_1$ WZW model with first two leading terms subtracted, rescaled by $T^{1/4}$, plotted up to $T=300$. If the Riemann hypothesis is true, then at large temperature, this function will remain bounded. However, if the Riemann hypothesis is false, at large temperatures the oscillations will grow in size and become unbounded (modulo the subtlety explained in footnote \ref{rhfoot}). In this plot, $\alpha \coloneqq 2\hat{E_1}(e^{2\pi i/3}) + \frac 3\pi (\gamma_E + \log(4\pi) + 24 \zeta'(-1)-2) \approx 0.975$ (see (\ref{eq:c2gen})). By fitting this plot with oscillating functions in $\log(T)$, we can numerically recover the first few nontrivial zeros of the zeta function. (A similar plot can be made for any $c>1$ CFT.)}
  \label{fig:SU31}
\end{figure}

For the case of the $SU(3)_1$ WZW model, we have $\rho = \sigma = e^{2\pi i/3}$. At large temperature, the last line of (\ref{eq:c2gen}) becomes non-perturbatively small. Therefore if we subtract the first two terms on the RHS of (\ref{eq:c2gen}) and go to large temperature, we should be able to probe the real part of the nontrivial zeros of the Riemann zeta function. Indeed, by evaluating (\ref{eq:su3scalars}) up to $T=300$, we numerically are able to recover the first few nontrivial zeros of the Riemann zeta function. We plot this in Fig. \ref{fig:SU31}. Of course for any 2d CFT we could make a similar plot using (\ref{eq:crossingggg}); here we picked this particular theory for concreteness.

We pause to note that we can only numerically go up to certain fixed temperature (e.g. $T=300$) because we only computed a finite number of terms in (\ref{eq:su3scalars}). Since the residue falls off exponentially in $\text{Im}(z_k)$, this means we only numerically test the Riemann hypothesis up to a fixed imaginary part. Since the Riemann hypothesis has already been checked up to imaginary part $3\times 10^{12}$ \cite{Platt}, we emphasize that our numerics are \emph{not} an independent check of the Riemann hypothesis.

However, it would be extremely interesting if there were a physical reason why the scalar partition function, with the first two leading terms subtracted off, had to scale as $T^{\frac c2 - \frac34}$. This would give a ``physics explanation" of the Riemann hypothesis. We leave this problem as an exercise to the reader.

\section{Future directions}
\label{sec:conclusion}

In this paper we have derived a crossing equation acting only on the scalar operators of certain 2d CFTs. Rather curiously the crossing equation is intimately related to the nontrivial zeros of the Riemann zeta function. This allows us to rephrase the Riemann hypothesis purely in terms of the growth of states of scalar operators of $U(1)^c$ CFTs. By applying clever choices of linear functionals, we are able to derive positive sum rules that the scalar operators must satisfy, which lead to nontrivial bounds on the lightest non-vacuum scalar operator in $U(1)^c$ CFTs. We discuss generalizations to theories with only Virasoro symmetry. There are various future directions that may be interesting to pursue.

\vspace{2mm}

\emph{Virasoro scalar crossing equation?}

\vspace{2mm}

In Sec. \ref{sec:virasoro} we derived a crossing equation acting on all operators for theories with Virasoro symmetry. In order to make the partition function square-integrable, we multiplied by a cusp form, namely $y^{c/2} |\eta(\tau)|^{2c}$, which led to the inclusion of all spins to the crossing equation. It would be nice if there exists a crossing equation that does not rely on this, and acts only on the scalar Virasoro primary operators. In order to derive such an equation (if it exists), it might be necessary to consider some generalization of harmonic analysis to allow for exponential divergences as $y\rightarrow \infty$.

In the end of Sec. \ref{sec:virasoro}, we guessed such a sum rule for Virasoro CFTs with $c<7$. It would be interesting to derive it more rigorously and somehow analytically continue the sum rule so it makes sense for arbitrary central charge. 

\vspace{2mm}

\emph{Crossing equation for spin $j$?}

\vspace{2mm}

In this paper we considered crossing equations acting on scalar operators of $U(1)^c$ CFTs (or ``fake scalars" for the case of Virasoro CFTs). This was largely to avoid the Maass cusp forms in the spectral decomposition (which lack scalars -- see (\ref{eq:massnoscalar})). It would be interesting if there were a generalization of our crossing equation to any fixed spin partition function. 

In fact, the techniques we studied almost immediately generalize to any spin $j \neq 0$ crossing equation. Let us denote $\mathcal{J}$ as the set of spin $j$ primary operators of a $U(1)^c$ CFT. The spin $j$ partition function is given by
\begin{align}
\sum_{\Delta \in \mathcal{J}} e^{-2\pi \Delta y} &= \frac{2\sigma_{c-1}(j) y^{\frac{1-c}2} K_{\frac{c-1}2}(2\pi j y)}{\Lambda\left(\frac c2\right)j^{\frac{c-1}2}} \nonumber \\ &~~~+ \frac1{2\pi i} \int_{\frac12-i\infty}^{\frac12+i\infty} ds \pi^{s-\frac c2} \Gamma\left(\frac c2-s\right)\mathcal{E}_{\frac c2-s}^c \frac{\sigma_{2s-1}(j)K_{s-\frac12}(2\pi j y) y^{\frac{1-c}{2}}}{\Lambda(s)j^{s-\frac12}} \nonumber \\
&~~~+ \sum_{n=1}^{\infty} \sum_{\epsilon=\pm} \frac{a_j^{(n,\pm)}(\hat Z, \nu_n^{\pm})}{2(\nu_n^{\pm},\nu_n^{\pm})}  y^{\frac{1-c}2} K_{i R_n^\pm}(2\pi j y).
\label{eq:spinjcrossing}
\end{align}
Unfortunately the last line in (\ref{eq:spinjcrossing}) seems very difficult to deal with analytically due to the sporadic nature of the Maass cusp form eigenvalues, but we can in fact do the integral in the second line using the same techniques as in Sec. \ref{sec:cross}. We move the contour of integration to the right, past $\text{Re}(s) = \frac c2$, so that we can expand the function $\mathcal{E}^c_{\frac{c}{2} + s - 1}$ in terms of the scalar primary operators and then change the order of the sum and integral. From the discussion in Appendix \ref{sec:crossder}, we know the only poles in $\mathcal{E}_{\frac c2-s}^c$ to the right of the contour occur at $s=\frac c2, \frac{1+z_n}2, \frac{1+z_n^*}2$ (see Fig. \ref{fig:polesc2}). The additional terms do not introduce any additional poles to the right of the contour. We thus get a crossing equation in terms of the spin $j$ operators on the LHS and the scalars on the RHS (as well as the cusp forms).  It may be interesting to analyze this equation further. 

\vspace{2mm}

\emph{Better bounds on $U(1)^c$ theories?}

\vspace{2mm}

In Table \ref{tab:table} our numerical bounds on the scalar operators of $U(1)^c$ theories are quite far from the average Narain scalar gap. For instance our numerical bounds seem to grow quadratically with $c$ instead of linearly. This may be an indication that our crossing equation is not strong enough to pinpoint the CFT with the largest scalar gap. It would be interesting if we could modify the set of crossing equations we consider to get better bounds. For instance we could include both our crossing and the ``traditional" modular invariance (or four-point function) crossing equations to see if we can get better bounds. Other avenues to explore may be to consider different functionals from the ones used in Sec. \ref{sec:functionalbohnanza} (for example not just considering $\varphi(t)$ in (\ref{eq:Phideff}) to be Gaussians) or somehow incorporate the residues at the nontrivial zeros of the zeta function into the crossing equation. It would also be nice to get numerical results for even $c$.

\vspace{2mm}

\emph{Four-point functions?}

\vspace{2mm}

There is a well-known relation between crossing symmetry acting on a four-point functions of four scalar operators and modular covariance. For four identical operators, under an appropriate coordinate transformation, the four-point function should be modular invariant. (For different operators, it will transform as some vector-valued modular function.) It would be interesting if one could derive a crossing equation on certain correlation functions where only a one-dimensional slice of operators are exchanged (e.g. only scalar operators are exchanged instead of operators of all spin). It would be especially interesting if this could generalize to higher dimensions. 

\vspace{2mm}

\emph{Applications to $\mathcal{N}=4$ SYM?}

\vspace{2mm}

Besides in 2d CFT, another natural place that modular invariance shows up in string theory is in $S$-duality of $\mathcal{N}=4$ super Yang-Mills theory. In \cite{Collier:2022emf} (see also \cite{Green:2014yxa}), harmonic analysis was used extensively to study various integrated correlators as a function of the complexified Yang-Mills coupling. It would be interesting if there were some sort of crossing equation acting only on the zero-instanton sector (but note that the pole structure of the overlap with Eisensteins is different because there is no notion of a ``scalar gap"; see \cite{Collier:2022emf} for discussions on this).

\section*{Acknowledgments}

We thank Scott Collier, Liam Fitzpatrick, Tom Hartman, Per Kraus, Yuya Kusuki, Ying-Hsuan Lin, Aike Liu, Alex Maloney, Dalimil Maz\'{a}\v{c}, Hirosi Ooguri, Danylo Radchenko, David Simmons-Duffin, Herman Verlinde, Yifan Wang, and Xi Yin for very helpful discussions. We are especially grateful to Scott Collier and Liam Fitzpatrick for numerous discussions on related topics; to Danylo Radchenko for telling us the functionals used around (\ref{eq:deffunc}); and to David Simmons-Duffin for enormous help on the numerics. We thank Liam Fitzpatrick and Tom Hartman for very helpful comments on a draft. NB is supported by the Sherman Fairchild Foundation. CHC is supported by Simons Foundation grant 488657 (Simons Collaboration on the Nonperturbative Bootstrap) and a DOE Early Career Award under grant no. DE-SC0019085. NB and CHC are grateful for the hospitality of the Bootstrap 2022 conference at the University of Porto, as well as the hospitality of United Airlines Flights 1990 and 144, during which part of this work was completed.

%\newpage

\appendix

\section{Pole structure of scalar crossing equation}
\label{sec:crossder}

In this appendix we will carefully derive the pole structure of the constrained Epstein zeta series $\mathcal{E}_s^c(\mu)$. Much of this analysis is in Sec. 3.2 of \cite{Benjamin:2021ygh}. Let us look at the scalar sector of (\ref{eq:spectraldecomp}):
\begin{align}
    y^{\frac c2}\left(1+ \sum_{\Delta \in \mathcal{S}} e^{-2\pi \Delta y}\right) &\coloneqq \int_{-1/2}^{1/2} dx \hat{Z}^c(\tau, \mu) \nonumber \\ &= y^{\frac c2} + \frac{\Lambda\left(\frac{c-1}2\right)}{\Lambda\left(\frac c2\right)} y^{1-\frac c2} + 3\pi^{-\frac c2} \Gamma\left(\frac c2 -1\right) \mathcal{E}^c_{\frac c2-1}(\mu) \nonumber \\ &~~~~+ \frac1{4\pi i}\int_{\frac12-i\infty}^{\frac12+i\infty} ds \pi^{s-\frac c2} \Gamma\left(\frac c2-s\right)\mathcal{E}_{\frac c2-s}^c(\mu) \left(y^s + \frac{\Lambda(s-\frac12)}{\Lambda(s)} y^{1-s}\right) \nonumber \\
    &= y^{\frac c2} + \frac{\Lambda\left(\frac{c-1}2\right)}{\Lambda\left(\frac c2\right)} y^{1-\frac c2} + 3\pi^{-\frac c2} \Gamma\left(\frac c2 -1\right) \mathcal{E}^c_{\frac c2-1}(\mu) \nonumber \\ &~~~~+ \frac1{2\pi i}\int_{\frac12-i\infty}^{\frac12+i\infty} ds \pi^{s-\frac c2} \Gamma\left(\frac c2-s\right)\mathcal{E}_{\frac c2-s}^c(\mu) y^s.
    \label{eq:scalargenC1}
\end{align}
In the last line of (\ref{eq:scalargenC1}), we used the functional equation that $\mathcal{E}^c_{\frac c2-s}(\mu)$ obeys, (\ref{eq:varepsfunc}).
We would like to move the contour in (\ref{eq:scalargenC1}) to the right, so we again need to classify all simple poles of the integrand with $\text{Re}(s) > \frac12$. As was argued in \cite{Benjamin:2021ygh}, there can only be poles we cross at $s=\frac c2$ and $s=\frac{1+z_n}2, \frac{1+z_n^*}2$. Let us review the argument. 

The idea is to take the inverse Laplace transform of (\ref{eq:scalargenC1}) to get the scalar density of states. We then integrate from $0$ to some number $\Delta$ (not including the vacuum), and demand that this vanishes for sufficiently small $\Delta$. This is due to the fact that the spectrum for a compact CFT is discrete, so in general there is a gap between the vacuum and first excited scalar state. A simple calculation shows the number of scalar operators (excluding the vacuum) below $\Delta$ is
\begin{align}
    N_0(\Delta) &= \frac{2\pi^c \zeta(c-1)\Delta^{c-1}}{(c-1)\Gamma(\frac c2)^2  \zeta(c)} + 12 \frac{2^{\frac c2} \Delta^{\frac c2} \mathcal{E}^c_{\frac c2-1}(\mu)}{c(c-2)} - \frac1{2\pi i}\int_{\frac12-i\infty}^{\frac12+i\infty} ds \frac{2^{\frac c2-s} \Delta^{\frac c2-s}\mathcal{E}^c_{\frac c2-s}(\mu)}{s-\frac c2} \nonumber \\
    &= \frac{2\pi^c \zeta(c-1)\Delta^{c-1}}{(c-1)\Gamma(\frac c2)^2  \zeta(c)} + 12 \frac{2^{\frac c2} \Delta^{\frac c2} \mathcal{E}^c_{\frac c2-1}(\mu)}{c(c-2)} \nonumber \\&~~~ + \frac1{2\pi i}\int_{\frac12-i\infty}^{\frac12+i\infty} ds \frac{2^{\frac c2-s} \Delta^{\frac c2-s} \Gamma(s) \Gamma(s+\frac c2 - 1) \zeta(2s) \mathcal{E}^c_{\frac c2+s-1}(\mu)}{\pi^{2s-\frac12}\Gamma(\frac c2 + 1 -s) \Gamma(s-\frac12)\zeta(2s-1)} 
    \label{eq:n0count}
\end{align}
Let us look at the last line of (\ref{eq:n0count}). In the limit of small $\Delta$, we must get $0$ for the integrated density of states, which means the integral must cancel the two power laws in $\Delta$ coming from the first two terms. In the integral in the last line of (\ref{eq:n0count}), we must close the contour to the left in the $s$-plane since $\Delta$ is small. This will tell us about the pole structure of $\mathcal{E}_{\frac c2+s-1}^c(\mu)$ for $\text{Re}(s) < \frac12$ (if we wanted to know the pole structure for $\text{Re}(s) > \frac12$ we would look at the first line of (\ref{eq:n0count}) and again close the contour to the left). In order to cancel the term that goes as $\Delta^{c-1}$, we need a pole at $s=1-\frac c2$. This comes from the term $\Gamma(s+\frac c2-1)$ in the numerator, with the others being finite. (Although the other gamma and zeta functions naively contribute poles and zeros for integer $c$, their combination is always finite.) Moreover in order for the residue to match, this fixes
\be
\mathcal{E}^c_{0}(\mu) = -1.
\label{eq:residue1}
\ee
We also need to cancel the second polynomial in (\ref{eq:n0count}). This comes from a pole at $s=0$, coming from the $\Gamma(s)$ term. We see the residue already matches the coefficient in (\ref{eq:n0count}) so we cannot constrain the value of $\mathcal{E}^c_{\frac c2-1}(\mu)$. Finally there can be no other poles with $\text{Re}(s) < \frac12$. Naively this tells us that $\mathcal{E}_{\frac c2 + s - 1}^c(\mu)$ cannot have any poles for $\text{Re}(s) < \frac12$, but this is too fast -- if the prefactor vanishes then $\mathcal{E}_{\frac c2 + s - 1}^c(\mu)$ can have a pole. The only zeros with $\text{Re}(s) < \frac12$ in the prefactor of the integrand are when $s = \frac{z_n}2, \frac{z_n^*}2$, coming from the $\zeta(2s)$ term. Thus $\mathcal{E}_{\frac c2 + s - 1}^c(\mu)$ can have a pole at $s=\frac{z_n}2, \frac{z_n^*}2$. We also know that $\mathcal{E}^c_{\frac c2+s-1}(\mu)$ must have zeros at $s=-\frac c2, -\frac c2-1, \cdots$ to cancel the poles from $\Gamma(s+\frac c2-1)$. 

Thus, looking at the integrand in (\ref{eq:scalargenC1}), we see the only poles to the right of the contour of integration are at $s=\frac c2$ and $s=1-\frac{z_n}2, 1-\frac{z_n^*}2$. (Using the functional equation for the zeta function, we can rewrite the last term as $s=\frac{1+z_n}2, \frac{1+z_n^*}2$.) The residue of the pole at $s=\frac c2$ is given in (\ref{eq:residue1}) and the residue at $s=\frac{1+z_n}2, \frac{1+z_n^*}2$ is just given by reading off the pole from integrating the partition function against an Eisenstein series at $s=\frac{z_n}2$ (see (\ref{eq:residuedelta})).

This fully reproduces the pole structure which we used to derive (\ref{eq:crossingfinal}).

\section{Functional action on crossing equation}
\label{sec:morefunc}

Let us consider the functional
\be
\mathcal{F}_k[h(t)]\coloneqq \int_0^\infty \frac{dt}t h(t) \sum_{m=1}^\infty e^{-\pi k t^2 m^2}.
\label{eq:funck}
\ee
We would like to apply this functional to each of the terms in (\ref{eq:cb}). To do so let us first compute:
\begin{align}
    \mathcal{F}_k[t^s e^{-\frac{A}{t^2}}] &\coloneqq \sum_{m=1}^\infty f_m^k(s,A) \nn\\
    \mathcal{F}_k[t^s e^{-B t^2} U(\alpha, \beta, B t^2)] &\coloneqq \sum_{m=1}^\infty g_m^{\alpha,\beta,k}(s,B),
    \label{eq:fstuffdef}
\end{align}
with $f_m^k(s,A)$ and $g_m^{\alpha,\beta,k}(s,B)$ defined as
\begin{align}
    f_m^k(s,A) &= \int_0^\infty dt~t^{s-1} e^{-\frac{A}{t^2} - \pi k t^2 m^2} \nn\\&= A^{s/4}k^{-s/4} m^{-s/2} \pi^{-s/4} K_{s/2}(2m\sqrt{k \pi A}) \nn\\
    g_m^{\alpha,\beta,k}(s,B) &= \int_0^\infty dt ~t^{s-1} e^{-B t^2 - \pi k t^2 m^2}U(\alpha,\beta,Bt^2) \nn\\
    &= \frac12\left(B+k\pi m^2\right)^{-s/2}\Bigg(\frac{\Gamma(\frac s2)\Gamma(1-\beta)~_2F_1(\alpha,\frac s2,\beta;\frac{B}{B+k \pi m^2})}{\Gamma(1+\alpha-\beta)} \nn\\&~~~~~+ \frac{\Gamma(\beta-1)(B+k \pi m^2)^{\beta-1}\Gamma(1-\beta+\frac s2)~_2F_1(1+\alpha-\beta,1-\beta+\frac s2,2-\beta; \frac{B}{B+k\pi m^2})}{\Gamma(\alpha)B^{\beta-1}}\Bigg).
    \label{eq:fgmdef}
\end{align}
Applying this term by term to (\ref{eq:cb}) we then get:
\begin{align}
     \sum_{\Delta\in\mathcal{S}}\sum_k \alpha_k \sum_{m=1}^\infty \Bigg[&4\pi \Delta f_m^k(-c, 2\pi \Delta) - c f_m^k(2-c, 2\pi \Delta) \nn\\& -\sum_{n=1}^\infty \frac{b(n) n^{c-2}}{\sqrt{\pi}}\Big((c-2) g_m^{-\frac12,\frac c2,k}(c,2\pi n^2 \Delta) - 4\pi n^2 \Delta g_m^{-\frac12, \frac c2, k}(c+2, 2\pi n^2 \Delta) \nn\\&~~~~~~~~~~~~~~~~~~~~~~~ + 2\pi n^2 \Delta g_m^{\frac12, \frac{c+2}2, k}(c+2, 2\pi n^2 \Delta)\Big)\Bigg] \nn\\
    &=  \zeta(c-1) \Gamma\left(\frac{c-1}2\right) \pi^{\frac{1-c}2} \sum_k \alpha_k \left(\frac c2 k^{\frac c2-1} + \left(\frac c2-1\right)k^{-\frac c2}\right).
    \label{eq:cb3}
\end{align}
The above equation is summed over an arbitrary choice of $k$'s and $\alpha_k$'s, subject to the constraints in (\ref{eq:sumconstraint}).

Remarkably, for odd $c \geq 3$, we can get closed form expressions for the sums over $m$ in (\ref{eq:cb3}). For $c=3$, (\ref{eq:cb3}) reduces to
\begin{align}
\sum_{\Delta \in \mathcal{S}}\sum_k& \alpha_k  \left[\frac{\sqrt 2 - e^{2\sqrt 2 \pi \sqrt{k \Delta}}(\sqrt 2 - 2\pi \sqrt{k\Delta})}{2(-1+e^{2\sqrt 2 \pi \sqrt{k \Delta}})^2 \sqrt \Delta} + \sum_{n=1}^\infty \frac{b(n) n \pi \cosh(\sqrt 2 n \pi \sqrt{\tfrac \Delta k})}{4 k^{\frac 32} \sinh^3(\sqrt 2 n \pi \sqrt{\tfrac \Delta k})}\right] \nn\\
&= \frac \pi 6 \sum_k \alpha_k \left(\frac 32 k^{\frac 12} + \frac 12 k^{-\frac 32}\right).
\end{align}

The sum over $n$ can be simplified to give
\begin{align}
\sum_{\Delta \in \mathcal{S}}\sum_k& \alpha_k  \left[\frac{\sqrt 2 - e^{2\sqrt 2 \pi \sqrt{k \Delta}}(\sqrt 2 - 2\pi \sqrt{k\Delta})}{2(-1+e^{2\sqrt 2 \pi \sqrt{k \Delta}})^2 \sqrt \Delta} + \frac{\pi e^{2\sqrt 2\pi \sqrt{\frac{\Delta}{k}}}}{(-1+e^{2\sqrt 2 \pi \sqrt{\frac{\Delta}{k}}})^2 k^{3/2}}\right] \nn\\
&= \frac \pi 6 \sum_k \alpha_k \left(\frac 32 k^{\frac 12} + \frac 12 k^{-\frac 32}\right).
\label{eq:c3sumrulewithks}
\end{align}

To simplify (\ref{eq:cb3}) for $c$ odd, $c \geq 5$, we first define the auxiliary functions:
\begin{align}
    \nu_1(c,0,m) &\coloneqq  \frac{(-1)^{\frac{c+1}2+m}\Gamma(c)}{(c-2)\Gamma(m+1)\Gamma(\frac{c+1}2-m)}, \nn\\ 
    \nu_1(c,n,m) &\coloneqq (-1)^{n+\frac{c-1}2+m} 2^{2n}(n+1)(c-\frac{(n+1)(n+2)}2)\Gamma(c-n-2)\nn\\&~~\times\sum_{i=0}^{m-1}\sum_{j=0}^{i+1} \frac{(-1)^{i+j}(i-j+1)^n}{\Gamma(j+1)\Gamma(n-j+2)\Gamma(m-i)\Gamma(\frac{c+3}2-n-m+i)}, ~~~~~~~~ n \neq 0 \nn\\
    \nu_2(c,n,m) &\coloneqq 2^{-2m}(c-3-2m)(c-1-2m)(c+1-2m)(c+3-2m)\Gamma(\frac{c-3}2+m)\nn\\
    &\times\sum_{i=0}^{\frac{c-1}2-m} \sum_{j=0}^{i+1} \frac{(-1)^{m+n+i+j}(i-j+1)^{\frac{c+1}2-m}}{\Gamma(j+1)\Gamma(\frac{c+5}2-m-j)\Gamma(n-i+1)\Gamma(m-n+i+1)} \nn\\
    \nu_3(c,n,m)&\coloneqq (-1)^{n+\frac{c-1}{2}+m} 2^{2n} (n+1)(n+2)(n+3)\Gamma(c-n-4) \nn\\&\times\sum_{i=0}^m \sum_{j=0}^{i+1}\frac{(-1)^{i+j}(i-j+1)^{n+2}}{\Gamma(j+1)\Gamma(n-j+4)\Gamma(m-i+1)\Gamma(\frac{c-3}2-n-m+i)}.
\end{align}
Then (\ref{eq:cb3}) becomes:
\begin{align}
    \sum_{\Delta\in \mathcal{S}}\sum_k \alpha_k \Bigg[&\frac1{2^{\frac{3c}2-3}\pi^{\frac{c-3}2}\Delta^{\frac{c-2}2}(-1+e^{2\sqrt 2 \pi \sqrt{k \Delta}})^{\frac{c+1}2}}\sum_{i=0}^{\frac{c-1}2} \sum_{j=0}^{\frac{c-1}2}\nu_1(c,i,j) e^{2\sqrt 2\pi\sqrt{k \Delta}j}(\sqrt{2k\Delta}\pi)^i \nn\\
    &+\sum_{n=1}^\infty \frac{b(n) e^{2\sqrt 2 \pi n \sqrt{\frac \Delta k}} n^{\frac{c-1}2} \pi}{2^{\frac{c+13}4}k^{\frac{c+3}4}\Delta^{\frac{c-3}4}(-1+e^{2\sqrt 2 \pi n \sqrt{\frac \Delta k}})^{\frac{c+3}2}}\sum_{i=0}^{\frac{c-1}2}\sum_{j=0}^{\frac{c-5}{2}} \frac{\nu_2(c,i,j) k^{\frac j2} e^{2i\sqrt{2}n \pi \sqrt{\frac \Delta k}}}{2^{\frac j 2}n^j \pi^j \Delta^{\frac j 2}}\Bigg] \nn\\
    &= \zeta(c-1) \Gamma\left(\frac{c-1}2\right) \pi^{\frac{1-c}2} \sum_k \alpha_k \left(\frac c2 k^{\frac c2-1} + \left(\frac c2-1\right)k^{-\frac c2}\right),~~~~~~~~ c~\text{odd}, ~c \geq 5.
    \label{eq:presimp}
\end{align}

The sum over $n$ in (\ref{eq:presimp}) can be done exactly, which gives:
\begin{align}
    \sum_{\Delta\in \mathcal{S}}\sum_k \alpha_k \Bigg[&\frac1{2^{\frac{3c}{2}-3}\pi^{\frac{c-3}2}\Delta^{\frac{c-2}2}(-1+e^{2\sqrt 2 \pi \sqrt{k \Delta}})^{\frac{c+1}2}}\sum_{n=0}^{\frac{c-1}2} \sum_{m=0}^{\frac{c-1}2}\nu_1(c,n,m) e^{2\sqrt 2\pi\sqrt{k \Delta}m}(\sqrt{2k\Delta}\pi)^n \nn\\
    &+\frac{e^{2\sqrt 2 \pi \sqrt{\frac \Delta k}}}{2^{\frac{3c}2-7}\pi^{\frac{c-7}2} k^2 \Delta^{\frac{c-4}2}(-1+e^{2\sqrt 2 \pi \sqrt{\frac{\Delta}k}})^{\frac{c+1}2}}\sum_{n=0}^{\frac{c-5}2}\sum_{m=0}^{\frac{c-3}2} \nu_3(c,n,m) e^{2\sqrt 2 \pi \sqrt{\frac{\Delta}{k}}m} (\sqrt{\tfrac{2\Delta}k}\pi)^n\Bigg] \nn\\
    &= \zeta(c-1) \Gamma\left(\frac{c-1}2\right) \pi^{\frac{1-c}2} \sum_k \alpha_k \left(\frac c2 k^{\frac c2-1} + \left(\frac c2-1\right)k^{-\frac c2}\right),~~~~~~~~ c~\text{odd}, ~c \geq 5.
    \label{eq:yay}
\end{align}
In the notation of (\ref{eq:2good2bfalse}), 
\be
\text{vac}(k) = -\zeta(c-1) \Gamma\left(\frac{c-1}2\right) \pi^{\frac{1-c}2} \left(\frac c2 k^{\frac c2-1} + \left(\frac c2-1\right)k^{-\frac c2}\right),
\label{eq:vac_k}
\ee
and $f(k, \Delta)$ is the term in the brackets of (\ref{eq:yay}). Using these definitions, an explicit calculation verifies the claim in (\ref{eq:thisguyvanishes}).

By examining the crossing equation (\ref{eq:yay}), we notice something interesting. Acting on the crossing equation with $k^{3/2} \partial_k$ gives us an expression that is antisymmetric under $k \leftrightarrow k^{-1}$. This gives us another way to rewrite the crossing equation that will turn out to work for all $c$ (not just odd $c$). Let us define the following function, using (\ref{eq:fgmdef})
\begin{align}
    h(c,k,\Delta) &\coloneqq \sum_{m=1}^{\infty} k^{3/2} \partial_k\left(4\pi\Delta f^k_m(-c,2\pi \Delta) - cf_m^k(2-c,2\pi \Delta)\right) \nonumber \\
    &=\sum_{m=1}^\infty \Bigg[2^{-\frac{c-4}4} k^{\frac{c+2}4} m^{\frac c2} \Delta^{-\frac{c-4}4} \pi c K_{\frac c2}(2\sqrt 2 m \pi \sqrt{k\Delta}) \nonumber \\&~~~~~~~~~ - 2^{-\frac{c+2}4} k^{\frac c4} m^{\frac{c-2}2} \Delta^{-\frac{c-2}4} (c(c-2)+8\pi^2 \Delta k m^2)K_{\frac{c-2}2}(2\sqrt2 m \pi \sqrt{k\Delta})\Bigg].
    \label{eq:hcgeneral}
\end{align}
The sum can be evaluated exactly in closed form for odd $c$, but exists and converges for any $c$. An equivalent formulation of our scalar crossing equation is:
\be
k^{3/2} \text{vac}'(k) + \sum_{\Delta \in \mathcal{S}} h(c,k,\Delta) - h(c,k^{-1},\Delta) = 0.
\label{eq:betterCross}
\ee
The sum rules used in (\ref{eq:basistouse}) are just the odd derivatives of $k$ (evaluated at $k=1$) of (\ref{eq:betterCross}). Finally, note that the term $k^{3/2}\text{vac}'(k)$ is simply the contribution of the vacuum state:
\begin{align}
k^{3/2} \text{vac}'(k) &=  \frac{\Lambda\left(\frac{c-1}2\right)c(c-2)}{4}\left(k^{\frac{1-c}{2}} - k^{\frac{c-1}{2}}\right) \nonumber \\ &= \lim_{\Delta\rightarrow 0} \left(h(c,k,\Delta) - h(c,k^{-1},\Delta)\right).
\end{align}

\section{$c=1$ and $c=2$ revisited}
\label{sec:c1}

In this appendix, we reconsider $U(1)^c$ theories at $c=1$ and $c=2$. Due to the pole structure of the function $\Lambda(s) \coloneqq \pi^{-s} \Gamma(s) \zeta(2s)$, the spectral decomposition and scalar crossing equation for these theories are slightly different than for $c>2$. This is related to the fact that the average genus $1$ partition function for $c=1$ and $c=2$ Narain CFTs diverges \cite{Afkhami-Jeddi:2020ezh, Maloney:2020nni}.
For both $c=1$ and $c=2$ we will first consider Narain CFTs, and then the potentially more general $U(1)^c$ theories.
We will use the notation 
\begin{align}
\tilde E_s(\tau) &\coloneqq \Lambda(s) E_s(\tau) \nonumber \\
&=  \Lambda(s) y^s + \Lambda(1-s) y^{1-s} + \sum_{j=1}^\infty \frac{4\sigma_{2s-1}(j)\sqrt{y} K_{s-\frac12}(2\pi j y)}{j^{s-\frac12}} \cos(2\pi j x)
\label{eq:estardef}
\end{align}
which we can see obeys $\tilde E_s(\tau) = \tilde E_{1-s}(\tau)$. This will make $s \leftrightarrow 1-s$ crossing manifestly invariant.

\subsection{$c=1$ reconsidered}
\label{sec:crossingc1subtlety}

The $c=1$ free boson is labeled by a radius $r$. In our convention, we will take the self-dual point (i.e. the $SU(2)_1$ WZW model) to be $r=1$ so that $T$-duality acts as $r\leftrightarrow r^{-1}$. The spectral decomposition of the reduced $c=1$ partition function is:
\begin{align}
    \hat{Z}^{c=1}(\tau, r) = r + r^{-1} + \frac1{4\pi i} \int_{\frac12-i\infty}^{\frac12+i\infty} ds 2 \tilde E_s(\tau) (r^{2s-1}+r^{1-2s}).
    \label{eq:c1decomp}
\end{align}
(See e.g. Sec. 3.1.1 of \cite{Benjamin:2021ygh} for derivation.) Notice that there are no Maass cusp forms in (\ref{eq:c1decomp}).

At $c=1$, our scalar crossing equation (\ref{eq:crossingfinal}) reduces to
\begin{align}
1 + \sum_{\Delta \in \mathcal{S}} e^{-2\pi \Delta y} &= -1 + \varepsilon_{c=1}(\mu) y^{-\frac12} + \sum_{k=1}^\infty \text{Re}\left(\delta_{k,c=1}(\mu)y^{\frac{z_k}2}\right) +\sum_{\Delta \in \mathcal S} \sum_{n=1}^{\infty} b(n)  \sqrt{\frac{2\Delta}y} e^{-\frac{2\pi n^2 \Delta}y},
\label{eq:c1full}
\end{align}
where as usual $\mu$ is some abstract coordinate that we include to emphasize which terms are theory-dependent.

Let us verify (\ref{eq:c1full}) for a free boson at radius $r$. From the explicit spectral decomposition (\ref{eq:c1decomp}), we know that the free boson at radius $r$ has $\varepsilon_{c=1}(\mu) = r+ r^{-1}$ and $\delta_{k,c=1}(\mu) = 0$. Moreover, the set of scalar operators $\mathcal{S}$ are simply operators with either zero momentum or zero winding number (recall at $c=1$, the spin of an operator is just the product of its momentum and winding number). Thus the set $\mathcal{S}$ is simply operators of dimension $\frac{m^2}{2r^2}$ and $\frac{m^2r^2}{2}$ for $m \in \mathbb{Z}_{>0}$, each with degeneracy $2$. Thus (\ref{eq:c1full}) reduces to
\be
2 y^{\frac12} + 2\sum_{m=1}^{\infty} (e^{-\pi m^2 r^2 y} + e^{-\pi m^2 r^{-2} y})y^{\frac12} = r + r^{-1} + 2\sum_{n=1}^\infty \sum_{m=1}^{\infty} b(n) m (r e^{-\frac{\pi n^2 m^2 r^2}y} + r^{-1} e^{-\frac{\pi n^2 m^2}{y r^2}}).
\ee
We can rewrite the RHS with new variables $m' = n m, n'= n$ (and dropping primes)
\be
2 y^{\frac12} + 2\sum_{m=1}^{\infty} (e^{-\pi m^2 r^2 y} + e^{-\pi m^2 r^{-2} y})y^{\frac12} = r + r^{-1} + 2\sum_{m=1}^\infty \sum_{n|m} b(n) \frac mn (r e^{-\frac{\pi m^2 r^2}y} + r^{-1} e^{-\frac{\pi m^2}{y r^2}}).
\ee
It can be shown from properties of the M\"obius $\mu$ function that
\be
\sum_{n|m} b(n) \frac mn = 1
\ee
for all $m$. Our crossing equation is then equivalent to
\be
2 y^{\frac12} + 2\sum_{m=1}^{\infty} (e^{-\pi m^2 r^2 y} + e^{-\pi m^2 r^{-2} y})y^{\frac12} = r + r^{-1} + 2\sum_{m=1}^\infty (r e^{-\frac{\pi m^2 r^2}y} + r^{-1} e^{-\frac{\pi m^2}{y r^2}}).
\ee
This simply follows from the modular transformation properties of the Jacobi theta functions. 

We would now like to derive a more general bound for $U(1)^c$ CFTs at $c=1$, without assuming the theory is a free boson compactified on a circle. This means we cannot assume that the $\delta_{k,c=1}$ terms in (\ref{eq:c1full}) necessarily vanish, so we need to apply the same functionals that we considered in Sec. \ref{sec:functionalbohnanza}. We first take a derivative with respect to $y$ to remove the $\varepsilon_{c=1}(\mu)$ term. This gives the analog of (\ref{eq:cb}):
\begin{align}
    \sum_{\Delta \in \mathcal{S}} &\left[\left(\frac{4\pi \Delta}t - t\right) e^{-\frac{2\pi \Delta}{t^2}} + \sum_{n=1}^{\infty} b(n) 4\pi\sqrt 2 n^2 t^4 \Delta^{\frac 32} e^{-2\pi\Delta n^2 t^2}\right] \nonumber \\ &=  2t + \sum_{k=1}^\infty \text{Re}\left(\delta_{k,c=1}(\mu) (z_k-2)t^{z_k}\right).
    \label{eq:c1b}
\end{align}
We next would like to apply the functional (\ref{eq:deffunc}) to (\ref{eq:c1b}), but there a slight subtlety. Recall that (\ref{eq:deffunc}) was designed so that 
\be
\mathcal{F}[t^s] \propto \zeta(s).
\ee
The last line of (\ref{eq:c1b}) has a term $2t$, which will naively give something proportional to $\zeta(1)$ which diverges. However, it can be shown the integral (\ref{eq:deffunc}) converges. The reason is that $M_{\varphi}(s)$ in (\ref{eq:zetavanish}) vanishes at $s=1$ which cancels the divergence of the zeta function. A careful analysis shows that if we choose $\varphi(t) = \sum_{i=1}^N \alpha_i e^{-\pi k_i t^2}$ (subject to the constraints (\ref{eq:sumconstraint})), then
\be
\mathcal{F}^{\varphi}[2t] = \sum_{i=1}^N \alpha_i \left(-\frac{\log k_i}{2\sqrt{k_i}}\right).
\ee
We then apply the same functional $\mathcal{F}^\varphi$ to the LHS of (\ref{eq:c1b}). This gives
\begin{align}
\sum_{k} \alpha_k \Bigg(\frac{\log k}{2\sqrt{k}} &+  \sum_{\Delta \in \mathcal{S}} \Big[\frac{\pi \sqrt\Delta(\text{coth}(\sqrt2 \pi \sqrt{k\Delta})-1)}{\sqrt2} + \frac{\log(1-e^{-2\sqrt2 \pi \sqrt{k\Delta}})}{2\sqrt k} \nonumber \\&~~~ + \sum_{n=1}^\infty b(n) \frac{\sqrt{k}~\text{coth}\left(\sqrt 2 n \pi \sqrt{\tfrac\Delta k}\right) + \sqrt 2 n \pi \sqrt \Delta~ \text{csch}^2\left(\sqrt 2 n \pi \sqrt{\tfrac\Delta k}\right)}{4kn} \Big]\Bigg) = 0.
\label{eq:c1divergesn}
\end{align}
The sum over $n$ in (\ref{eq:c1divergesn}) formally diverges but we can replace $\text{coth}\left(\sqrt 2 n \pi \sqrt{\tfrac\Delta k}\right)$ with $\text{coth}\left(\sqrt 2 n \pi \sqrt{\tfrac\Delta k}\right)-1$ since the term we add is multiplied by $0$ from (\ref{eq:sumconstraint}). This gives the following convergent sum rule:
\begin{align}
\sum_{k} &\alpha_k \Bigg(\frac{\log k}{2\sqrt{k}} +  \sum_{\Delta \in \mathcal{S}} \Big[\frac{\pi \sqrt\Delta(\text{coth}(\sqrt2 \pi \sqrt{k\Delta})-1)}{\sqrt2} + \frac{\log(1-e^{-2\sqrt2 \pi \sqrt{k\Delta}})}{2\sqrt k} \nonumber \\&~~~ + \sum_{n=1}^\infty b(n) \frac{\sqrt{k}~\left(\text{coth}\left(\sqrt 2 n \pi \sqrt{\tfrac\Delta k}\right)-1\right) + \sqrt 2 n \pi \sqrt \Delta~ \text{csch}^2\left(\sqrt 2 n \pi \sqrt{\tfrac\Delta k}\right)}{4kn} \Big]\Bigg) = 0.
\label{eq:c1nodivergesn}
\end{align}
The sum over $n$ can be done exactly to give:
\begin{align}
\sum_{k} &\alpha_k \Bigg(\frac{\log k}{2\sqrt{k}} +  \sum_{\Delta \in \mathcal{S}} \Big[\frac{\pi \sqrt\Delta(\text{coth}(\sqrt2 \pi \sqrt{k\Delta})-1)}{\sqrt2} + \frac{\log(1-e^{-2\sqrt2 \pi \sqrt{k\Delta}})}{2\sqrt k} \nonumber \\&~~~ + \frac{\pi \sqrt\Delta(\text{coth}(\sqrt2 \pi \sqrt{\tfrac{\Delta}k})-1)}{\sqrt2k} - \frac{\log(1-e^{-2\sqrt2 \pi \sqrt{\tfrac{\Delta}{k}}})}{2\sqrt k}
\Big]\Bigg) = 0.
\label{eq:c1convergesn}
\end{align}
Again from the same arguments as used to derive (\ref{eq:2good2bfalse}) we know that the term in parenthesis in (\ref{eq:c1convergesn}) must be $c_0 + c_1 k^{-1/2}$ for some (theory-dependent) constants $c_0, c_1$. Moreover we see after evaluating $\partial_k|_{k=1}$ on each term, that $c_1=-1$. Therefore we can write our crossing equation as
\be
\text{vac}^{(n)}(1) + \sum_{\Delta \in \mathcal{S}} (\partial_k)^n f(k, \Delta)|_{k=1} = 0, ~~~n \geq 2, ~~n ~\text{even},
\label{eq:c1dervseven}
\ee
with
\begin{align}
    \text{vac}(k) &= \frac{2+\log k}{2\sqrt k} \nonumber \\
    f(k, \Delta) &= \frac{\pi \sqrt\Delta(\text{coth}(\sqrt2 \pi \sqrt{k\Delta})-1)}{\sqrt2} + \frac{\log(1-e^{-2\sqrt2 \pi \sqrt{k\Delta}})}{2\sqrt k} \nonumber \\&~~~ + \frac{\pi \sqrt\Delta(\text{coth}(\sqrt2 \pi \sqrt{\tfrac{\Delta}k})-1)}{\sqrt2k} - \frac{\log(1-e^{-2\sqrt2 \pi \sqrt{\tfrac{\Delta}{k}}})}{2\sqrt k}. 
\end{align}
Note that the equations (\ref{eq:c1dervseven}) are indeed equivalent to derivatives (with respect to $k$, evaluated at $k=1$) of (\ref{eq:betterCross}) at $c=1$.

\subsection{$c=2$ reconsidered}
\label{sec:crossderc2}

The $c=2$ free boson is labeled by a metric and $B$ field, which gives four real moduli in total. These can be repackaged into two elements of the upper half plane as \cite{Dijkgraaf:1987jt}:
\begin{align}
    \rho = B + i\sqrt{\text{det}~G}, ~~ \sigma=\frac{G_{12}}{G_{11}} + i \frac{\sqrt{\text{det}~G}}{G_{11}}.
\end{align}
$T$-duality acts as two independent elements of $SL(2,\mathbb Z)$ acting on $\rho$ and $\sigma$ in the usual way. In terms of these coordinates, the spectral decomposition of the reduced $c=2$ partition function is:
\begin{align}
    \hat{Z}^{c=2}(\tau,\rho,\sigma) &= \hat{E_1}(\tau) + \hat{E_1}(\rho) + \hat{E_1}(\sigma) - \frac{3}{\pi}\(4 - \gamma_E - 3\log(4\pi) - 48\zeta'(-1)\) \nonumber \\
    &~~~~+ \frac{1}{4\pi i} \int_{\frac12 - i\infty}^{\frac12 + i\infty} ds \frac{\tilde E_s(\tau) \tilde E_s(\rho) \tilde E_s(\sigma)}{\Lambda(s) \Lambda(1-s)} \nonumber \\
    &~~~~+ 8\sum_{n=1}^{\infty} \frac{\nu^+_n(\tau) \nu^+_n(\rho) \nu^+_n(\sigma)}{(\nu^+_n, \nu^+_n)} - 8 i\sum_{n=1}^{\infty} \frac{\nu^-_n(\tau) \nu^-_n(\rho) \nu^-_n(\sigma)}{(\nu^-_n, \nu^-_n)}.
    \label{eq:c2spec}
\end{align}
(See e.g. Sec. 3.1.2 of \cite{Benjamin:2021ygh} for derivation.) In (\ref{eq:c2spec}), the function $\hat{E_1}$ is defined as
\begin{align}
\hat{E_1}(\tau) &\coloneqq \lim_{s\rightarrow1} E_s(\tau) - \frac{3/\pi}{s-1} \nonumber \\
&= y - \frac3\pi \log y + \frac{6}{\pi}\left(1-12\zeta'(-1) - \log 4\pi\right) + \sum_{j=1}^\infty \frac{12\sigma_1(j) e^{-2\pi j y} \cos(2\pi j x)}{j}.
\label{eq:e1hat}
\end{align} 

Let us derive the scalar crossing equation at $c=2$. We first assume the theory is a Narain CFT. As usual let us denote the set of scalar operators under the $U(1)^2$ chiral algebra excluding the vacuum, as $\mathcal{S}$. (Of course, $\mathcal{S}$ depends on the moduli of the theory, which for $c=2$ we denote by $\rho, \sigma$, but we will suppress that.) The partition function of these scalars is given by
\begin{align}
    y\left(1 +  \sum_{\Delta \in \mathcal{S}} e^{-2\pi \Delta y}\right) &\coloneqq \int_{-1/2}^{1/2} dx \hat{Z}^{c=2}(\tau, \rho, \sigma) \nonumber \\ &= y - \frac3\pi \log y  + \hat{E_1}(\rho) + \hat{E_1}(\sigma)  + \frac{3}{\pi}\left(-2+24\zeta'(-1) + \gamma_E + \log 4\pi\right)\nonumber \\
    &~~~~+ \frac{1}{4\pi i} \int_{\frac12 - i\infty}^{\frac12 + i\infty} ds \frac{(\Lambda(s) y^s + \Lambda(1-s) y^{1-s}) \tilde E_s(\rho) \tilde E_s(\sigma)}{\Lambda(s) \Lambda(1-s)} \nonumber \\ 
    &= y - \frac3\pi \log y   + \hat{E_1}(\rho) + \hat{E_1}(\sigma) + \frac{3}{\pi}\left(-2+24\zeta'(-1) + \gamma_E + \log 4\pi\right) \nonumber \\
    &~~~~+ \frac{1}{2\pi i} \int_{\frac12 - i\infty}^{\frac12 + i\infty} ds \frac{y^s \tilde E_s(\rho) \tilde E_s(\sigma)}{\Lambda(1-s)}.
    \label{eq:c2specscalar}
\end{align}
Let us move the contour in $s$ to the right past all the poles. The function $\tilde E_s$ has simple poles at $s=0,1$ (which can be seen from (\ref{eq:estardef})). Moreover, $\Lambda(1-s) = \pi^{\frac12-s} \Gamma(s-\frac12)\zeta(2s-1)$ has zeros whenever $2s-1$ is a nontrivial zero of the Riemann zeta function. Thus the integrand has simple poles that we cross at $s=1, \frac{1 + z_n}2, \frac{1+z_n^*}2$, where $z_n$ is a nontrivial zero of the Riemann zeta function (with positive imaginary part).\footnote{The double pole at $s=1$ in the numerator of the integrand becomes a simple pole when canceled by the simple pole at $s=1$ in the denominator. There is also a pole at $s=0$, but since we move the contour to the right we can ignore it.} A picture of the pole structure is given in Fig. \ref{fig:polesc2} (where we move the pole at $s=\frac c2$ to $s=1$).

We then get the equation:
\begin{align}
 y \left(1+\sum_{\Delta \in \mathcal{S}} e^{-2\pi \Delta y}\right) &= - \frac3\pi \log y + \frac{3}{\pi}\left(-2+24\zeta'(-1) + \gamma_E + \log 4\pi\right) + \hat{E_1}(\rho) + \hat{E_1}(\sigma) \nonumber \\
&~~~~+  \sum_{k=1}^{\infty} \text{Re}\left(\frac{4\pi^{\frac{z_k}{2}}  \Lambda(\frac{1+z_k}2)^2 E_{\frac{1+z_k}2}(\rho)  E_{\frac{1+z_k}2}(\sigma)}{2\Gamma\left(\frac{z_k}2\right)\zeta'(z_k)} y^{\frac{1+z_k}2}\right) \nonumber \\
&~~~~ + \frac1{2\pi i}\int_{\gamma-i\infty}^{\gamma+i\infty} ds \frac{y^s \tilde E_s(\rho) \tilde E_s(\sigma)}{\Lambda(1-s)},
\label{eq:naivecrossingc2}
\end{align}
where $\gamma > \frac c2 = 1$.
This integral is a special case of the one studied in (\ref{eq:thisIsU}), which can be done exactly to give us:
\begin{align}
 1+\sum_{\Delta \in \mathcal{S}} e^{-2\pi \Delta y} &= - \frac3\pi \frac{\log y}y + \frac{\frac{3}{\pi}\left(-2+24\zeta'(-1) + \gamma_E + \log 4\pi\right) + \hat{E_1}(\rho) + \hat{E_1}(\sigma)}y \nonumber \\
&~~~~+  \sum_{k=1}^{\infty} \text{Re}\left(\frac{4\pi^{\frac{z_k}{2}}  \Lambda(\frac{1+z_k}2)^2 E_{\frac{1+z_k}2}(\rho)  E_{\frac{1+z_k}2}(\sigma)}{2\Gamma\left(\frac{z_k}2\right)\zeta'(z_k)} y^{\frac{-1+z_k}2}\right) \nonumber \\
&~~~~ + \frac{1}{y\sqrt \pi} \sum_{\Delta \in \mathcal{S}} \sum_{n=1}^\infty b(n) U\left(-\frac12, 1, \frac{2\pi \Delta n^2}y\right) e^{-\frac{2\pi \Delta n^2}y}.
\label{eq:Exactcrossingc2}
\end{align}
The sum over $k$ in (\ref{eq:Exactcrossingc2}) falls off exponentially in $k$ so the sum is indeed convergent.

The generalization to any $U(1)^2$ CFT at $c=2$ is straightforward. We again need to subtract $\hat{E_1}(\tau)$ to render the reduced partition function square-integrable, and a gap to the first excited state constrains the poles we cross in $s$ to only be at $s=1, \frac{1+z_n}2, \frac{1+z_n^*}2$ (see Appendix \ref{sec:crossder}). Finally, the same arguments as in Sec \ref{sec:cross} let us compute the non-perturbative corrections at high temperature to get:
\begin{align}
 1+\sum_{\Delta \in \mathcal{S}} e^{-2\pi \Delta y} &= - \frac3\pi \frac{\log y}y + \frac{\varepsilon_{c=2}(\mu)}y +  \sum_{k=1}^{\infty} \text{Re}\left(\delta_{k,c=2} ~ y^{\frac{-1+z_k}2}\right) \nonumber \\
&~~~~ + \frac{1}{y\sqrt \pi} \sum_{\Delta \in \mathcal{S}} \sum_{n=1}^\infty b(n) U\left(-\frac12, 1, \frac{2\pi \Delta n^2}y\right) e^{-\frac{2\pi \Delta n^2}y}.
\label{eq:Exactcrossingc2noNarain}
\end{align}

\bibliographystyle{JHEP}
\bibliography{ScalarRH}
\end{document}